# Low-energy electron microscopy as a tool for analysis of self-assembled molecular layers on surfaces.

*Jan Čechal,*[1,2*] *Pavel Procházka*[1]

[1]CEITEC – Central European Institute of Technology, Brno University of Technology, Purkyňova 123, 612 00 Brno, Czech Republic.

[2]Institute of Physical Engineering, Brno University of Technology, Technická 2896/2, 616 69 Brno, Czech Republic.

AUTHOR INFORMATION

**Corresponding Author**

* E-mail: cechal@fme.vutbr.cz (J. Č.)

ABSTRACT

Low-energy electron microscopy (LEEM) is a surface science method that works primarily in the UHV environment. It provides information complementary to the other established techniques: it extends the limited view of scanning probe microscopies from nanometers to micrometers and measurement time down to tens of milliseconds, enabling to visualize the changes during sample treatment, e.g., annealing, deposition, and gas or light exposure. From the point of structural analysis, it allows the measurement of diffraction patterns from an area of diameter below 200 nm and imaging of phase distribution on the surfaces either through dark-filed imaging or LEEM-I(V) fingerprinting. The advanced modes provide local angle-resolved photoelectron spectra and surface potential distribution. In this review, we aim to describe the utilization of LEEM to study self-assembled molecular structures on solid surfaces. We present the LEEM instrumentation and analysis of measured data in a tutorial way to provide the necessary background knowledge to enter the field. In the second part, we summarize the knowledge obtained by LEEM for several selected systems, which points to the strength of LEEM in understanding the self-assembled molecular systems and its synergy with other surface science techniques.

# 1. Introduction

Low-energy electron microscopy (LEEM) is a surface science technique primarily operating under ultrahigh vacuum (UHV) conditions. LEEM employs low-energy electrons (typically 0–50 eV) to image the samples in real-space at a mesoscopic scale. The measurements can be performed in real-time (typically 0.2 s per frame) during sample annealing, irradiation, deposition of new material, or exposure to a specific gas, thus allowing monitoring of changes occurring in response to the treatment. The LEEM instrument inherently comprises the low-energy electron diffraction (LEED) mode, so we can deduce the structure of observed materials with high precision from the measured diffraction patterns. Placing apertures into the electron beam enables advanced LEEM modes. Spatially restricting the incoming electron beam allows for measuring the diffraction on areas with a diameter of ~185 nm (depending on the LEEM instrument used). Selecting only a particular diffracted beam (blocking all others) either largely enhances the contrast (bright-field imaging) or images only those parts of the surface having the given structure (dark-field imaging). The later mode enables mapping of the spatial distribution of phases or symmetry-equivalent domains on the surface, which is usually termed phase analysis in materials science. When combined with a suitable UV radiation source, the LEEM instrument can operate as a photoemission electron microscope (PEEM) and an energy filter to measure local angle-resolved photoemission spectroscopic data or energy-filtered images.

In this review, we will focus on the applications of LEEM on self-assembled molecular phases on solid surfaces, specifically, the organic molecules forming extended molecular phases through non-covalent interactions. In this respect, LEEM gives us mesoscale information on the molecular phases, especially their formation and transformation during growth and thermal annealing. The true capabilities of LEEM are much enhanced when combined with other methods, especially within a single UHV system, which allows measuring the same sample using multiple techniques. Here, the real-time capability is advantageous; it enables us to stop the measurement and do either scanning tunneling microscopy (STM) or X-ray photoelectron spectroscopy (XPS) when a particular change takes place, e.g., a new phase is formed. The mesoscale view confirms the uniformity of the sample, which is essential for the area-integrated XPS measurements that provide chemical information necessary for the interpretation of observed changes. STM helps to relate the distinct contrast in the LEEM images and the diffraction patterns with the structure of the molecular phases.

Low-energy electron microscopy was invented by Ernst Bauer [1,2]. Over the years, the technique matured, and a new design by Ruud Tromp was developed [3]. These two designs are now commercially available by Elmitec and SPECS, respectively. The technique was thoroughly described in several review papers [2,4,5] and book chapters [6–9]. In addition, applications of LEEM were reviewed several times with a focus on a specific topic developed by the authors [10–17]. Following application areas were reviewed: (i) clean metals and their alloys [6,7], growth of thin films [6,7] and reactions [7] on metal surfaces; (ii) phase transitions on Si surfaces including step and phase boundary fluctuations and surface roughening and faceting [7,10], surface phenomena on wide band semiconductors [7]; (iii) oxide and nitride surfaces [7], nucleation, growth, and transformation of oxides on transition and rare-earth metal substrates [13]; (iv) graphene on SiC and metal surfaces, transferred graphene [11], and other two dimensional van der Waals materials [12]; (v) organic thin films with focus on linear conjugated molecules pentacene and anthracene [14], diindenoperylene [15], and a metal-free single molecular magnet, NitPyn [15]; (vi) application of intensity–voltage I(V) spectroscopy [8,16] with focus on structural phenomena, diffusion, alloying and chemical reactions in thin films on metal and oxide substrates [16]; (vii) imaging magnetic structures with spin-polarized LEEM [6,17]; (viii) local angle-resolved photoelectron spectroscopy and angle resolved reflected electron spectroscopy for mapping band structure of occupied and unoccupied bands and electron energy loss spectroscopy [8]; (ix) local potentiometry and work-function mapping [8]; and (x) radiation effects with LEEM [8]. Related technique, photoemission electron microscopy (PEEM) [18], is a powerful method enabling time-resolved measurements [18] and monitoring reactions on surfaces [19,20].

Here, we review the applications of LEEM for analyzing self-assembled molecular layers on surfaces and recent developments in this field enabled by this technique. We intend to provide the application tutorial, and hence, we will not describe everything measured in detail but rather demonstrate LEEM capabilities on selected organic systems. As mentioned above, LEEM is stronger when combined with other methods, preferentially performed on the same sample without breaking the vacuum. In this respect, we will discuss the findings alongside the STM and XPS results.

In the following, we will first describe the LEEM instrumentation (Section 2), then discuss the measured data and how to understand them (Section 3), and finally, summarize the results obtained by LEEM on selected molecular systems (Section 4).

## 2. LEEM Instrumentation.

This section briefly describes the LEEM instrument. Its design was described and reviewed several times [2,3,6–8], allowing us to be brief. We will focus our description on the design by Ruud Tromp [3,8,21], commercially available by SPECS (SPECS FE-LEEM P90), and briefly point to differences to the design by Ernst Bauer [2,7] offered by Elmitec. There is also a newcomer in the field designed by Wen-Xin Tang and made available by Suzhou AISTech.

LEEM is an imaging technique that employs a parallel beam of electrons in the way the classical optical microscope uses light. It images the entire irradiated view field at the same time. To study solid surfaces, we need to measure electrons reflected from the surface; as a consequence, the LEEM system has a bent optical axis. Next, the electrons should be of low energy to achieve high surface sensitivity and high resolution. These quantities are defined by the mean free path of electrons, which decreases with decreasing electron energy from thousands of eV typical in standard electron microscopes to its lower limit occurring below 50 eV. It would be challenging to design the electron optics for these low energies. Therefore, throughout the optics, the electron energy is 15 keV, and the electrons are decelerated to a few eV just in front of the sample. This requires that the sample has the same potential as the electron emitter, typically −15 kV, and limits samples to conductive ones.

## 2.1 Optical elements

LEEM consists of the electron source (emitter), electron optics (condensor, objective, and imaging columns), sample stage, apertures, and sample prism, whose functionality is detailed in the following subsection. Simplified schematics of the SPECS FE-LEEM P90 instrument with all the main components is shown in Figure 2.1a.

*Gun and condensor column.* The electron source is a cold-field-emission electron gun kept at a high potential of −15 kV, while the electron columns are at ground potential. Electrons extracted from the emitter have an energy spread of ~0.25 eV, which limits the image resolution to 3–4 nm, assuming that no other factors compromise/enhance it [3]. Overall, a lateral resolution of 5 nm is achievable by a standard instrument [3]; with an aberration corrector, the limiting resolution is decreased to 2 nm [22,23]. A magnified electron source image (virtual source) is projected to the sample prism by gun and condenser lenses.

*Sample prism.* The sample prism is a beam splitter; it spatially separates the condensor and projector optics [3]. It first deflects the electron beam by 90 ° toward the objective lens and the electrons reflected from the sample by 90 ° toward the projector column. The prism also serves as an energy filter, allowing the LEEM system to measure angle-resolved energy spectra and energy-filtered images when combined with UV light sources [24]. The Elmitec design is based on a 60° deflector prism.

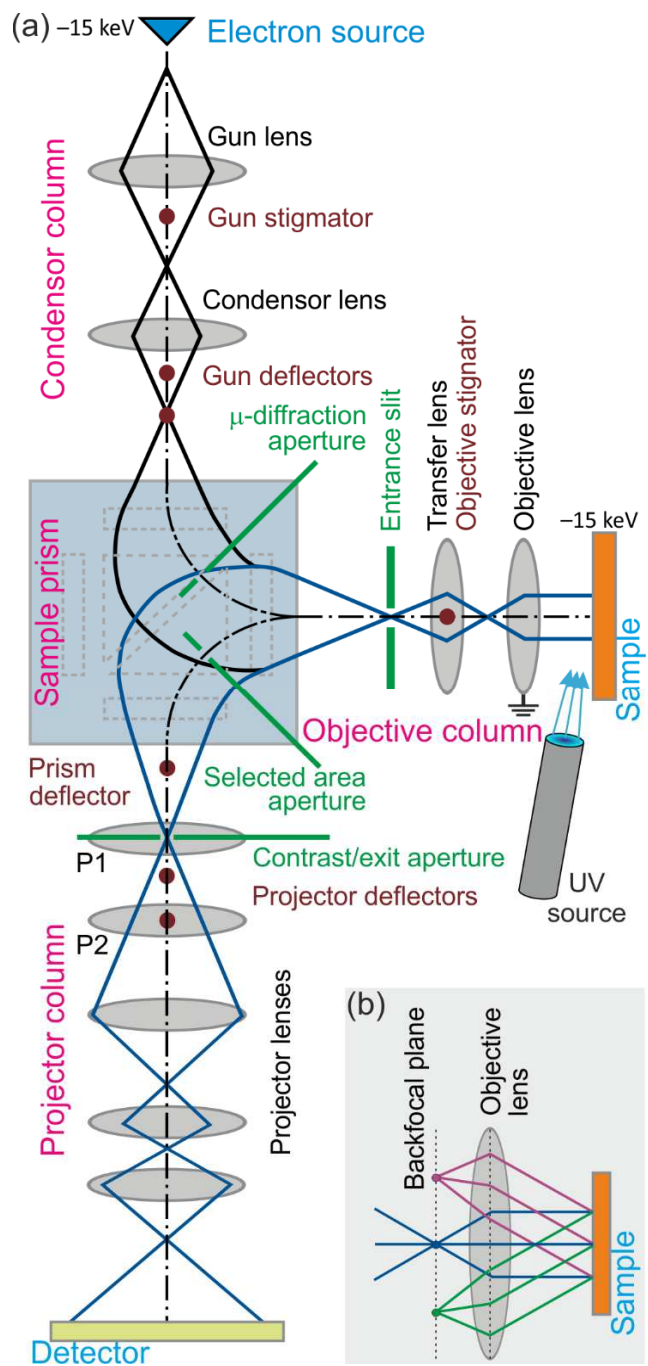

**Figure 2.1.** IBM/SPECS LEEM Setup. (a) The electrons emitted from a cold-field-emission electron gun are projected by a condensor column to the sample prism that bends the beam by 90° to the objective column, which projects a parallel electron beam on the sample. The electric field between the grounded objective lens and sample held at negative potential decelerates the electrons

to low energies. The electrons reflected from the sample are accelerated back to the transfer energy and focused into the prism, where the beam is bent by 90 ° to the projector column, which magnifies the image on the detector. In addition to the lenses, there are two sets of stigmators and three sets of deflectors, which adjust the beam position and tilt angle. (b) The magnified view around the sample shows the formation of diffraction in the back focal plane of the objective lens. Figure (a) adapted from ref. [24] with permission from SPECS Surface Nano Analysis GmbH, manufacturer of the instrument.

*Objective column and sample.* The sample is maintained at a potential close to that of the electron emitter, i.e., typically −15 kV, while the electron optics is at ground potential. An electrostatic immersion objective lens, therefore, places the sample in a strong uniform electrostatic field (~10 MV/m), decelerating the electrons to the required low energies of a few eV [3]. The high electric field on the sample put strict requirements on the sample and sample holder, which should be conducive and relatively flat without sharp features to avoid sparking. While the field may seem strong, it is three orders of magnitude lower than in STM, where a non-uniform local electric field of several GV/m is between the sample and the STM tip.

The objective lens provides a parallel beam of electrons landing on the sample surface and focuses the reflected electrons back into the prism. The electron beams reflected in the same direction are focused on the same points in the back focal plane of the objective lens, as shown in Figure 2.1b. Hence, the diffraction pattern is formed in the back focal plane while the real-space image is formed further away [8].

*Projector column.* The prism deflects the reflected beam to the projector column, which is dedicated to magnifying and transferring the image onto the detector. The back focal plane of the objective lens featuring the diffraction pattern is projected to the center of the projector lens P1. It is where an aperture can be inserted to select diffraction spots for bright- and dark-field imaging. Depending on the excitation of the projector lens P2, either real- or reciprocal-space image is projected on the detector. Turning the P2 lens on/off enables fast switching between imaging the sample in real space and recording the diffraction pattern. Following projector lenses magnify the image or diffraction pattern on the detector.

The detector traditionally comprises a microchannel plate, which amplifies the electron signal. It hits the fluorescent screen, transforming it into a light signal the CCD camera detects. New detectors, e.g., direct electron cameras, offer new possibilities and may replace traditional detectors [25,26].

## 2.2 Apertures

Several apertures of various sizes throughout the LEEM system regulate the beam. They increase the image contrast, enable the selection of the part of the real or reciprocal space to be imaged, or allow ARPES or energy-filtered imaging.

*µ-diffraction aperture.* µ-diffraction aperture in the sample prism restricts the area irradiated by the electron beam on the sample from ~7×15 µm down to a circle with a diameter down to 185 nm. This allows us to obtain diffraction patterns from much smaller areas.

*Contrast aperture.* The contrast aperture is placed at the center of the projector lens P1, where the prism refocuses the diffraction pattern. It allows us to select a particular beam to be imaged, i.e., the central beam for bright-field imaging or any other diffraction spot for dark-field imaging.

*Entrance slit.* The entrance slit restricts the angular distribution of the electrons in one dimension to a small band enabling recording angle-resolved energy spectra and, in combination with the contrast aperture acquisition of energy-filtered images.

*Selected area aperture.* The selected area aperture controls the size of the reflected electron beam that leaves the prism. For LEEM measurement, this aperture is usually open. For PEEM experiments (especially spectroscopy), it is used to adjust the viewing area.

## 2.3 Other sources

The design of the objective lens close to the sample allows the implementation of additional excitation sources or deposition cells at grazing incidence angles (70 ° with respect to the surface normal). Typically, UV sources like Hg arc lamp (~4.9 eV), He discharge lamp (21.22 or 40.81 eV), or deposition cells are installed here, as shown in Figure 2.1a. Note that at the angle of 70 °, the deposition rate decreases to 1/3 of the nominal rate.

## 3. What information does LEEM provide about molecular layers?

LEEM provides several imaging modes. Without apertures, we obtain a real space image and a diffraction pattern from an irradiated area of ~7×15 μm. Inserting the *μ-diffraction aperture* restricts the irradiated area to a circle with a diameter down to 185 nm. In this way, we get the diffraction pattern only from the irradiated area; this mode is called μ-diffraction. Placing the *contrast aperture* and selecting only the central (0 0) spot for real-space imaging increases the contrast; this mode is called bright-field (BF) imaging. Selecting any higher-order spot enables us to image only areas that contribute to the selected diffraction spot, i.e., has the particular structure; this mode is called dark-field imaging. In the following, we explain these imaging modes in detail, with a focus on self-assembled molecular systems. We will unconventionally start the topic with diffraction, which allows a more straightforward explanation of bright-field and dark-filed imaging.

## 3.1 Understanding diffraction.

*"Diffraction speaks clearly."*

### 3.1.1 Fundamentals of diffraction

The diffraction on a 2D lattice is a part of each surface science textbook [27–30]. Here, we only briefly summarize the fundamentals. In three dimensions, the diffraction condition for elastic scattering is given by the Laue equations $\mathbf{K} = \mathbf{G}$, in which $\mathbf{K} = \mathbf{k} - \mathbf{k_0}$ is scattering vector, $\mathbf{k_0}$ and $\mathbf{k}$ are the wave vectors of the incident and scattered wave, respectively, and $\mathbf{G}$ is a vector of the reciprocal space lattice [31]. At the surface, the bulk periodicity is truncated, and the Laue conditions reduce to two equations regarding the components of the scattering and wave vectors parallel to the surface: $\mathbf{K}_\parallel = \mathbf{k}_\parallel - \mathbf{k_{0\parallel}} = \mathbf{G}_\parallel$, where $\mathbf{G}_\parallel$ is a 2D reciprocal lattice vector associated with the surface. The graphical solution to diffraction condition, so called Ewald's construction, is shown in Figure 3.1a.

A two-dimensional solid surface is described by a 2D Bravais lattice $\mathbf{R} = m\mathbf{a_1} + n\mathbf{a_2}$ with primitive basis vectors $\mathbf{a_1}$ and $\mathbf{a_2}$. A 2D reciprocal lattice in reciprocal space is associated with the Bravais lattice in real space. The primitive basis vectors $\mathbf{a_1^*}$ and $\mathbf{a_2^*}$ of the 2D reciprocal lattice $\mathbf{G} = h\mathbf{a_1^*} + k\mathbf{a_2^*}$ are defined according to the orthogonality relation $\mathbf{a_i} \cdot \mathbf{a_j^*} = 2\pi\delta_{ij}$ with $\delta_{ij}$ being Kronecker delta and $i, j = 1, 2$. With $\mathbf{n}$ being a unit vector perpendicular to the surface, the orthogonality relation gives the following expressions for reciprocal lattice vectors:

$$\mathbf{a_1^*} = 2\pi \frac{\mathbf{a_2} \times \mathbf{n}}{|\mathbf{a_1} \times \mathbf{a_2}|}; \quad \mathbf{a_2^*} = 2\pi \frac{\mathbf{n} \times \mathbf{a_1}}{|\mathbf{a_1} \times \mathbf{a_2}|} \qquad (1).$$

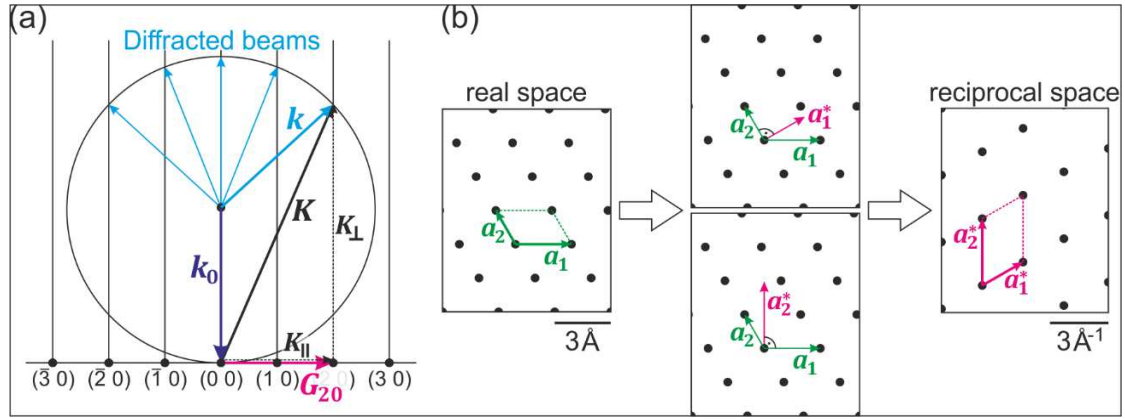

**Figure 3.1.** Diffraction and reciprocal lattice. (a) Ewald construction provides a graphical solution to diffraction conditions. The endpoint of the incident wave vector $\boldsymbol{k_0}$ is placed to any point of the reciprocal lattice; in the diffraction experiment, $\boldsymbol{k_0}$ is typically perpendicular to the surface, so it is drawn here. The Ewald sphere (projected on the circle here) represents the condition of elastic scattering, i.e., $|\boldsymbol{k}| = |\boldsymbol{k_0}|$. The Laue diffraction conditions express the conservation of in-plane momentum as $\boldsymbol{K}_\parallel = \boldsymbol{k}_\parallel = \boldsymbol{G}_\parallel$; this is illustrated for $\boldsymbol{G}_\parallel = \boldsymbol{G}_{20}$. As there is no periodicity in the out-of-plane direction, any amount of momentum can be transferred in this direction, which is expressed as crystal truncation rods perpendicular to the surface, each placed at the reciprocal lattice point. The $\boldsymbol{k}$ vector of diffracted beams then points from the center of the Ewald sphere to the intersection of the Ewald sphere with truncation rods, fulfilling thus all given conditions. (b) Translation from real to reciprocal space. Note that in the middle panels, we show vectors in two different spaces; the real space vectors have units of Å and reciprocal space vectors Å$^{-1}$.

The basic rules for transition from real to reciprocal space are (see Figure 3.1b): (i) what is larger in real space becomes smaller in reciprocal space; for example, for a square lattice, the equations (1) reduce to $\boldsymbol{a_1^*} = \frac{2\pi}{a_1}$ and $\boldsymbol{a_2^*} = \frac{2\pi}{a_2}$, i.e., the length of reciprocal vectors is proportional to the reciprocal value of real space vectors. (ii) $\boldsymbol{a_1^*}$ is perpendicular to $\boldsymbol{a_2}$ and $\boldsymbol{a_2^*}$ is perpendicular to $\boldsymbol{a_1}$, see vector multiplication in the numerator of (1). (iii) The real space lattice is described by primitive basis vectors $\boldsymbol{a_1}$ and $\boldsymbol{a_2}$, these are translated to primitive reciprocal lattice vectors $\boldsymbol{a_1^*}$ and $\boldsymbol{a_2^*}$, the linear combination of $\boldsymbol{a_1^*}$ and $\boldsymbol{a_2^*}$ gives a reciprocal lattice. Of course, the real space lattice contains double, triple,… periodicities, but these are not translated to the reciprocal lattice.

In a diffraction experiment, each diffracted beam corresponds to a reciprocal lattice vector $\boldsymbol{G_\parallel} = h\boldsymbol{a_1^*} + k\boldsymbol{a_2^*}$, where $h$ and $k$ are integers. Hence, the diffraction pattern corresponds to the image of a reciprocal lattice except for spot intensities. Employing the dynamical diffraction theory allows the calculation of spot intensities as a function of energy. By comparison with measured data, it enables a very precise determination of the surface structure, i.e., the positions of atoms in the few topmost layers.

Compared to the standard LEED setup with a spherical screen with an electron gun mounted in its center, reciprocal space imaging in LEEM has several advantages: the central (0 0) spot is visible, and the diffraction spot positions do not change with energy (if a flat surface is measured) [7]. On the other hand, LEEM comprises magnetic lenses and a prism; hence, the diffraction image may be distorted, making the quantitative measurements more difficult.

### 3.1.2 Reading the diffraction pattern

In the diffraction pattern of our interest, there are contributions from at least two lattices: the substrate and molecular layer on top. The diffraction pattern is the image of the reciprocal lattice except for spot intensities. Hence, it contains all the information about the real-space lattices except for their mutual position. If more lattices are present, the associated diffraction patterns are superimposed and aligned to the central spot; i.e., the mutual shifts of real-space lattices are not translated to the diffraction pattern.

We will present the analysis of a diffraction pattern taking the Ag(001) surface featuring the semi-deprotonated β phase of BDA molecules as an example [32]; this molecular system is detailed in Section 5.1. The diffraction pattern measured from the large area of 7×15 μm$^2$ is given in Figure 3.2a. At first sight, it is quite complex and may be challenging to understand. This hints that the diffraction pattern likely comprises several contributions. Indeed, employing μ-diffraction aperture, we can distinguish four distinct μ-diffraction patterns presented in Figure 3.2b. Their color-coded composition, given in Figure 3.2c, presents the complete large-area diffraction pattern in Figure 3.2a, confirming that we have a complete set of μ-diffraction patterns.

When the domains are smaller than the analyzed area provided by the smallest μ-diffraction aperture, moving the μ-diffraction aperture over the sample surface results in brightening some spots while dimming others as the contribution from different domains within the analyzed area changes. In this case, all spots associated with a given domain brighten/darken simultaneously.

Considering the substrate $p4mm$ symmetry sketched in Figure 3.2d, i.e., four-fold rotational symmetry and two sets of mirror planes inclined by 45 °, we expect the presence of 2 or 4 symmetry equivalent molecular domains: 4 in general case and 2 in high symmetry case when the molecules within the molecular-phase domains align with principal surface directions. The substrate symmetry operations applied to one of the μ-diffraction patterns in Figure 3.2b produce the other three, proving that the presence of 4 distinct molecular domains is due to the symmetry of the surface. Models of these four domains are given in Figure 3.2e; in all cases, the position of BDA molecules with respect to the surface is the same. Hence, the number of domains usually reflects the symmetry of the surface; e.g., for a two-fold symmetric molecular layer on a hexagonal surface with $p6mm$ symmetry, we expect either 6 or 3 symmetry-equivalent domains.

Figure 3.2f shows the decomposition of the μ-diffraction pattern to its components: substrate, molecular, and moiré spots. Spots associated with molecular phases are usually closer to the central (0 0) spot than the substrate ones, as the lattice parameter of molecular phases is larger than for the substrate. The best way to identify the substrate spots is to record the growth of the molecular layer starting from the clean surface. Alternatively, the substrate spots keep their position in all microdiffraction patterns, and they are usually intense for sub-monolayer coverages.

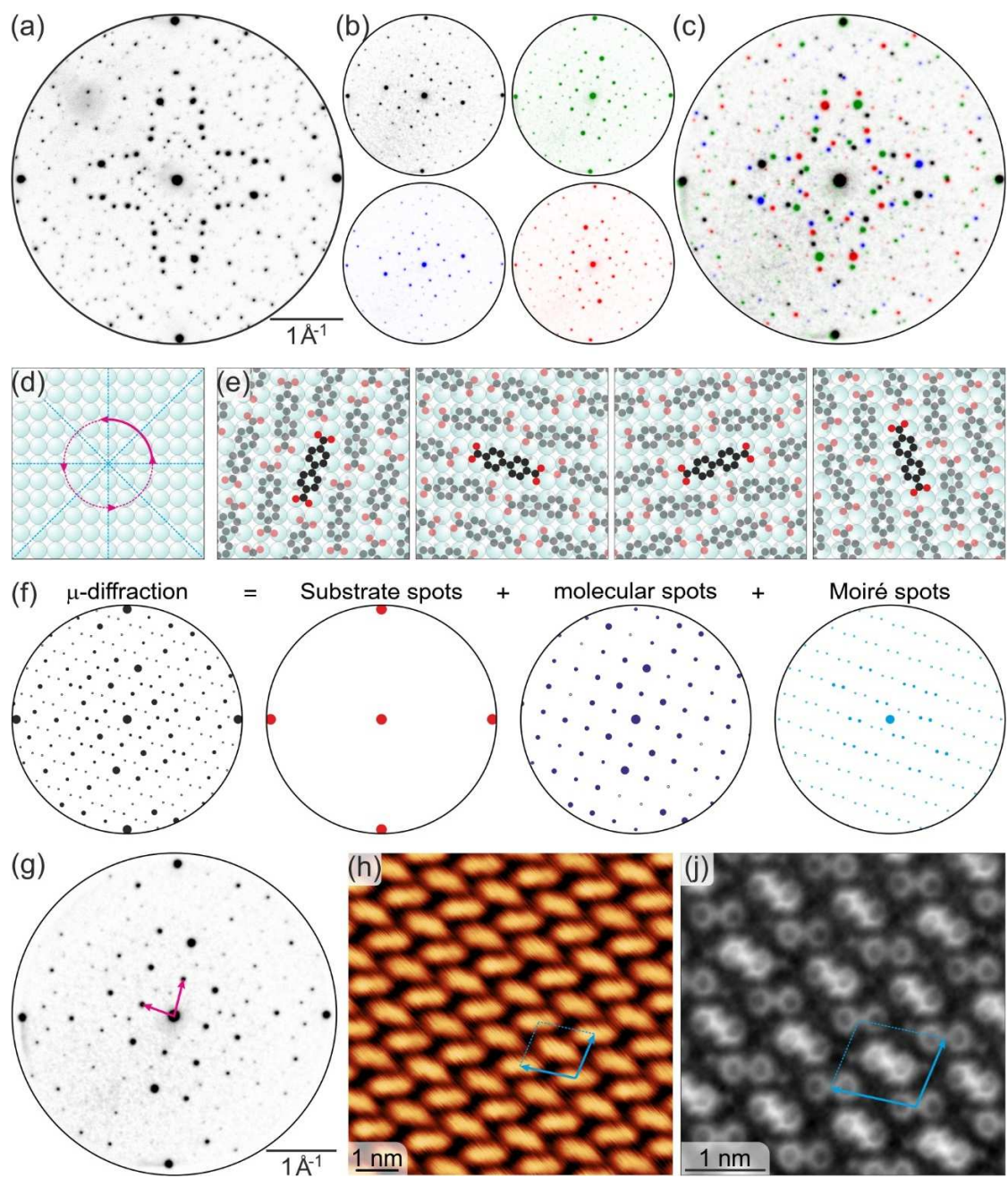

**Figure 3.2.** Diffraction analysis of a molecular structure. Data for the BDA β phase on the Ag(001) substrate measured with a primary energy of 20 eV are used throughout the image. (a) Measured large-area diffraction pattern. (b) µ-diffraction patterns of four BDA domains and (c) their color-coded composition. (d) Symmetry elements of Ag(001) substrate $p4mm$ point group: the four-fold rotation symmetry is sketched by an arrow and mirror planes by dashed lines. (e) Four

symmetry-equivalent domains obtained by applying substrate symmetry operations. (f) Decomposition of the μ-diffraction pattern into substrate, molecular, and moiré spots. (g) Magnified μ-diffraction pattern of one BDA domain and associated (h) STM and (j) nc-AFM images. The reciprocal and real space unit vectors are marked by arrows. Parts of the figure were adapted with permission from the ref. [32]. Copyright 2020 American Chemical Society.

The diffraction spots associated with molecular domains reflect the reciprocal lattice; therefore, they provide quantitative information about the real-space unit cell, i.e., its size and orientation with respect to the substrate. However, we should approach the direct measurement of distances with care, as the diffraction patterns measured by LEEM are often distorted [6]; still, the measurement provides reasonable estimates of lattice size and orientation, especially when done relatively to the substrate. To determine the molecular structure, we need to put the basis (i.e., the structural objects, usually one molecule) to the lattice points. Here, the real-space image of the structure provided, e.g., by STM or AFM, is of huge help.

The magnified μ-diffraction pattern of one BDA domain and the associated STM and non-contact AFM images are shown in Figure 3.2g, h, and j. The STM image in Figure 3.2h portrays all the molecules as oval-shaped protrusions with two distinct orientations. It enables us to draw an apparent unit cell in real space, which reasonably matches the reciprocal unit cell from diffraction. In many cases, the measured diffraction pattern points to a larger unit cell than the one we would directly draw to the STM image, mainly due to minute alterations in the position/shape of certain molecules as shown in the nc-AFM image in Figure 3.2j, which reveals slight bending of BDA molecule in the center of the unit cell.

The analysis of the diffraction pattern can go much further, overcome the distortions in the image, and determine the unit cell very precisely. The presence of moiré spots in the diffraction pattern (Figure 3.2f) is a sign that the BDA unit cell is still commensurate with the substrate but with a much larger periodicity than a brief look at diffraction, STM, and nc-AFM would suggest. Only two intense moiré spots are visible in each unit cell in the measured µ-diffraction pattern given in Figure 3.2g. The position of these spots helps to identify the full unit cell. Moiré spots appear when two lattices with different periodicities overlap, making a new, larger periodicity. Large organic molecules may be relatively weakly bound to the substrate by van der Waals interactions, and due to their size, they do not simply match the underlaying substrate lattice. Instead, the molecular superstructure is usually commensurate with the substrate with a periodicity of several apparent molecular unit cells.

To make the analysis of the very complex molecular diffraction patterns featuring multiple domains feasible, we can employ the local congruence method, which takes advantage of this complexity. It is based on a comparison of motifs formed by spots from different symmetry-equivalent domains, as shown in Figure 3.3a. The mutual position of diffraction spots originating from different domains is very sensitive to the unit cell shape, size, and orientation. To apply the method, we model a diffraction pattern and adjust it until it matches the measured one. Figure 3.3b shows a measured diffraction pattern with two regions magnified. The best-fit model matches the experiment perfectly. If the unit cell within the model is slightly changed (the last two images in Figure 3.3b), then the diffraction patterns are significantly distinct, showing a very high sensitivity of this approach for unit cell determination. The software tool ProLEED Studio is of great help for this analysis as it allows us to directly modify the reciprocal space and translate the changes to real space [33].

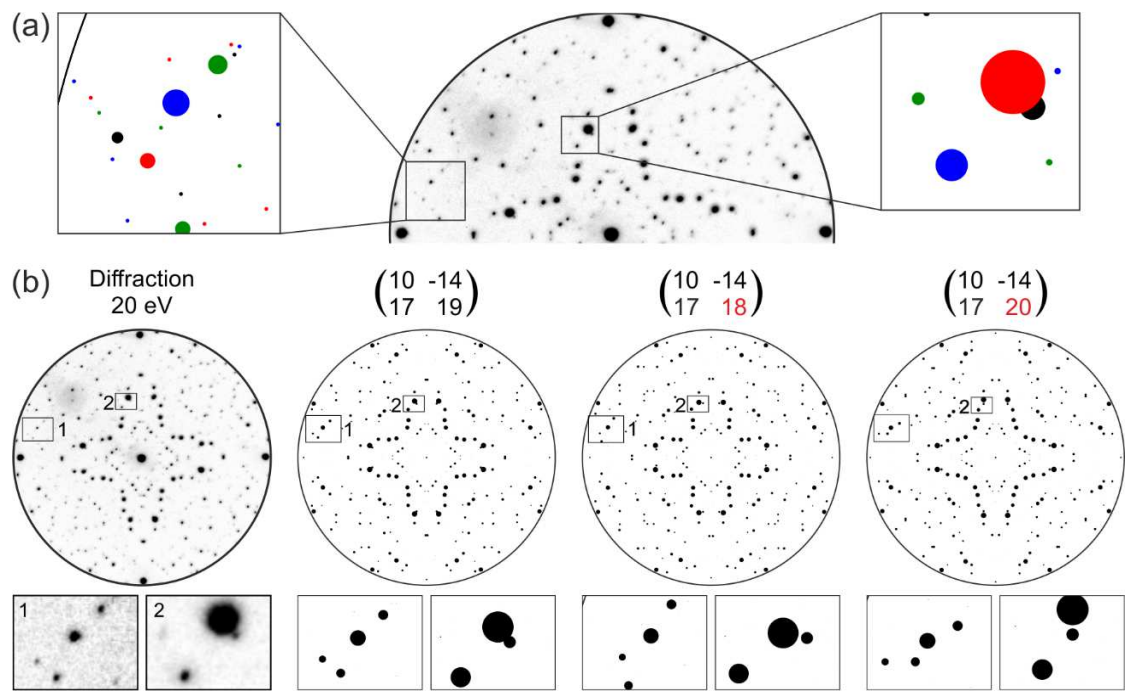

**Figure 3.3.** Local congruence analysis of a diffraction pattern. In the local congruence approach, we compare selected patterns formed by diffraction spots from distinct symmetry-equivalent molecular domains. (a) Magnified view of a selected pattern in which the color of the spots is associated with distinct domains shown in Figures 3.2b and c. (b) Measured diffraction pattern (left), the calculated diffraction pattern of the best-fit model (moiré spots are not shown), and two calculated diffraction patterns associated with models with slightly different unit cells. Two selected areas are magnified below the full diffraction images. Parts of the figure were adapted with permission from the ref. [32], copyright 2020 American Chemical Society; and ref. [33], copyright 2024 The Authors, published by International Union of Crystallography under CC-BY 4.0 license.

### 3.1.3 Spot intensities

For low-energy electrons, the multiple scattering is significant; hence, the kinematic theory of diffraction fails to describe the spot intensities correctly. Intensity-vs-energy I(V) curve analysis presents one of the most precise experimental methods for determining the surface structure with pm precision [31,34,35]. It is based on a comparison of measured I(V) curves in a broad energy range (30 – 400 eV) for several diffraction spots with calculated intensities employing dynamical diffraction theory [31,34,35]. This type of analysis was almost abandoned due to its complexity and time demands, as well as the rise in the popularity of ab initio calculations, with the belief that it can fully replace LEED experiments. Fortunately, recent efforts revitalize the I(V) analysis and make it easy to use via the ViPErLEED package [36,37]. Separately, spot profile analysis could provide in-depth information on studied material [8], but this goes beyond the scope of this review.

While LEEM presents several advantages, the accessible energy range is inferior to dedicated LEED instruments [6], and the I(V) curve analysis by LEEM is uncommon even with respect to inorganic systems. Concerning molecular layers, classical I(V) modeling would be too demanding, considering the huge size of unit cells compared with inorganic systems. Hence, so far, there have been no applications of I(V) curve analysis to determine the structure in molecular systems. The more frequent approach is to record the I(V) curve for the central (0 0) beam for each pixel of the LEEM image [6,8], which will be discussed as a part of bright field imaging in the next section.

Separately, spot profile analysis in LEED also presents a powerful technique of structure analysis, but similarly to the I(V) curve analysis, distortion in LEEM makes it largely underperforming the SPA-LEED instruments [6]. While the dynamic theory treatment is necessary for the correct calculation of the spot intensities, there are few cases in which spot intensities give reliable information even without the full theoretical treatment. These will be discussed in the following paragraphs.

If the 2D structure comprises glide symmetry planes, certain spots are systematically absent, also referred to as kinematically forbidden or missing spots [38,39]. Glide symmetry is the combination of mirror reflection and a shift by half of a unit cell in the direction parallel to the glide plane, which is highlighted in Figure 3.4d. For surface-confined structures, the glide plane is defined by vector $\boldsymbol{R}$ in the surface plane and normal vector $\boldsymbol{n}$. If the $\boldsymbol{k}$ vector of scattered wave lies in a glide plane, only the even-order spots are visible, and the odd-order spots are missing. For example, if the glide plane is parallel to the $y$-axis, in the set of $(0\ n)$ diffraction spots, all the spots for which $n$ is odd are missing. Figure 3.4a shows a diffraction pattern associated with $(4\sqrt{2}\times 4\sqrt{2})R45°$ superstructure of BDA on Cu(001), where certain posts are apparently missing. The STM image in Figure 3.4b and a model in Figure 3.4c show molecular structure featuring alternating orientation of BDA molecules, which gives rise to two sets of glide symmetry planes. These planes lie in the direction of BDA unit cell vectors, as marked by dashed lines in Figure 3.4b. The existence of glide planes means that every odd spot in the direction of BDA reciprocal unit cell vectors should be absent, specifically …$(\bar{5}\ 0)$, $(\bar{3}\ 0)$, $(\bar{1}\ 0)$, $(1\ 0)$, $(3\ 0)$, $(5\ 0)$,… spots for horizontal glide plane and …$(0\ \bar{5})$, $(0\ \bar{3})$, $(0\ \bar{1})$, $(0\ 1)$, $(0\ 3)$, $(0\ 5)$,… for vertical glide plane. Indeed, these spots are missing, as highlighted by light blue circles in Figure 3.4a. Similar absences are present for every second BDA-associated spot in these directions around each substrate spot, as marked by red circles in Figure 3.4a.

Breaking the glide symmetry, e.g., by slight distortion of molecules, makes the supposedly absent spots visible, but for slight distortions, they are usually still weak. For example, the BDA δ-phase on Ag(001) surface [40] shows weak spots instead of absent ones. The apparent glide symmetry is distorted by different bending of the end-groups towards the surface, which is apparent by their slightly different contrast in STM images [40,41].

Apart from glide symmetry, specific arrangements of identical atoms in the unit cell may produce a lattice of missing spots at all energies due to vanishing structure factors in the limit of kinematic scattering [42]. We also note that if a centered unit cell is used, then the spots of the type $\frac{1}{2}h + \frac{1}{2}k$ with $h + k$ odd are missing, but this is just a consequence of use of non-primitive basis vectors [42]; this essentially means that avoiding centered unit cells makes the diffraction analysis easier.

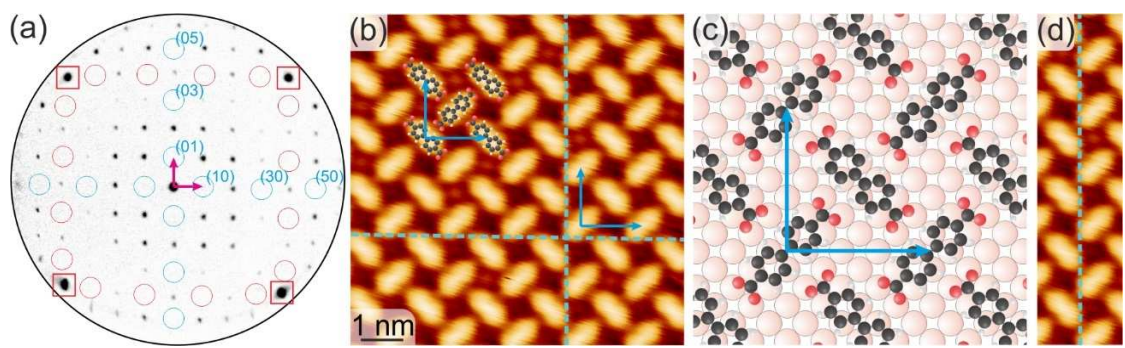

**Figure 3.4.** Systematic absence of diffraction spots. (a) Diffraction pattern of BDA on Cu(001); first order substrate spots are marked by red squares and the missing spots by circles. (b) STM image of BDA on Cu(001) and (c) associated model shows the position and orientation of molecules. The unit cell is marked by cyan arrows. The projection of two glide planes on the surface is marked by dashed lines. (d) Cut from STM image in (b) highlighting the glide symmetry. Figure adapted with permission from ref. [43]. Copyright 2018 American Chemical Society.

The second example concerns the distinct spot intensities from self-assemblies with distinct chirality. Figures 3.5a and b show STM images in which the honeycomb networks comprise molecules with distinct chirality, which is reflected in the intensities of spots in the µ-diffraction patterns of the domains with different chiralities as highlighted by cyan ellipses in Figures 3.5 c and d [44]. Typically, the lattices with distinct chirality show a significant mutual rotation [45–47], making them easily distinguishable. The changes in spot intensities for the same lattices formed by Co and phthalocyanines (CoPc and $F_{16}CuPc$) featuring distinct orientations of individual molecules were pointed out by Antczak and explained as the expression of molecule structure factor [48]. Hence, the spot intensities may be used as fingerprints to distinguish similar lattices with distinct structures even without full calculations. Finally, we mention that there could be changes in spot intensities if the second molecular layer is positioned non-symmetrically with respect to the first layer, which was employed for visualization of molecular stacking via dark-field measurements [49].

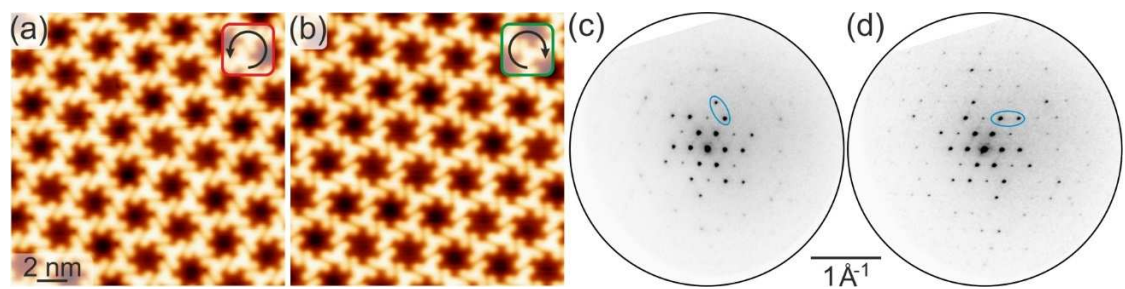

**Figure 3.5.** Chirality translated to spot intensities. (a) and (b) STM images of two domains comprising propeller-like molecules with distinct chirality forming the honeycomb lattice. (c) and (d) µ-diffraction patterns measured on domains with distinct chirality. Distinct spot intensities associated with a distinct chirality of domains are highlighted by cyan ellipses. Figure adapted with permission from ref. [44]. Copyright 2025 The Authors, published by Wiley-VCH GmbH under CC BY-NC 4.0 license.

## 3.1.4 Linear features in diffraction patterns

Some diffraction patterns contain linear features in addition to sharp diffraction spots. The appearance of lines can be a result of overlapping diffraction patterns that originate from many small domains with multiple mutual orientations [50], or the lines reflect the breaking of the periodicity of lattices by domain walls and boundaries [34]. Figure 3.6a shows the diffraction pattern associated with Fe-tetracyanobenzene metal-organic framework on the Au(111) surface. In addition to diffraction spots, it comprises lines connecting them. The STM in Figures 3.6 b and c shows the corresponding structure in real space; instead of long-range order, we identify preferential structural motifs that can be assembled together in many ways. However, diffraction streaks do not arise from any disorder but from the breaks in periodic lattices, in this case by "assembling together" structural motifs "in many ways". This is further demonstrated by our exemplar molecular system featuring BDA molecules. Figure 3.6d portrays the as-deposited BDA molecular phase on the Ag(001) surface ($\alpha$ phase), whose diffraction pattern contains streaks. Due to balancing molecule-molecule and molecule-substrate interactions, the BDA is not periodically spaced within the molecular chains [40]; however, this does not have a major effect on the resulting diffraction pattern [51]. This is demonstrated in Figure 3.6f, where the molecule-shaped objects were randomly displaced from their original positions (Figure 3.6d). The FFT shows a decrease in spot intensities but no shifts of blurring of individual spots. The random displacement of scatterers from the ideal position is an integral part of the dynamical diffraction calculations; consistently with the FFT simulations, this fundamental theory also describes the general decrease in spot intensities with increasing random disorder but not spot blurring or splitting. However, once the twin domain boundaries were introduced, clear streaks in the FFT signal appeared, as documented in Figure 3.6g. The diffraction streaks in reciprocal space are always perpendicular to the orientation domain boundaries in real space.

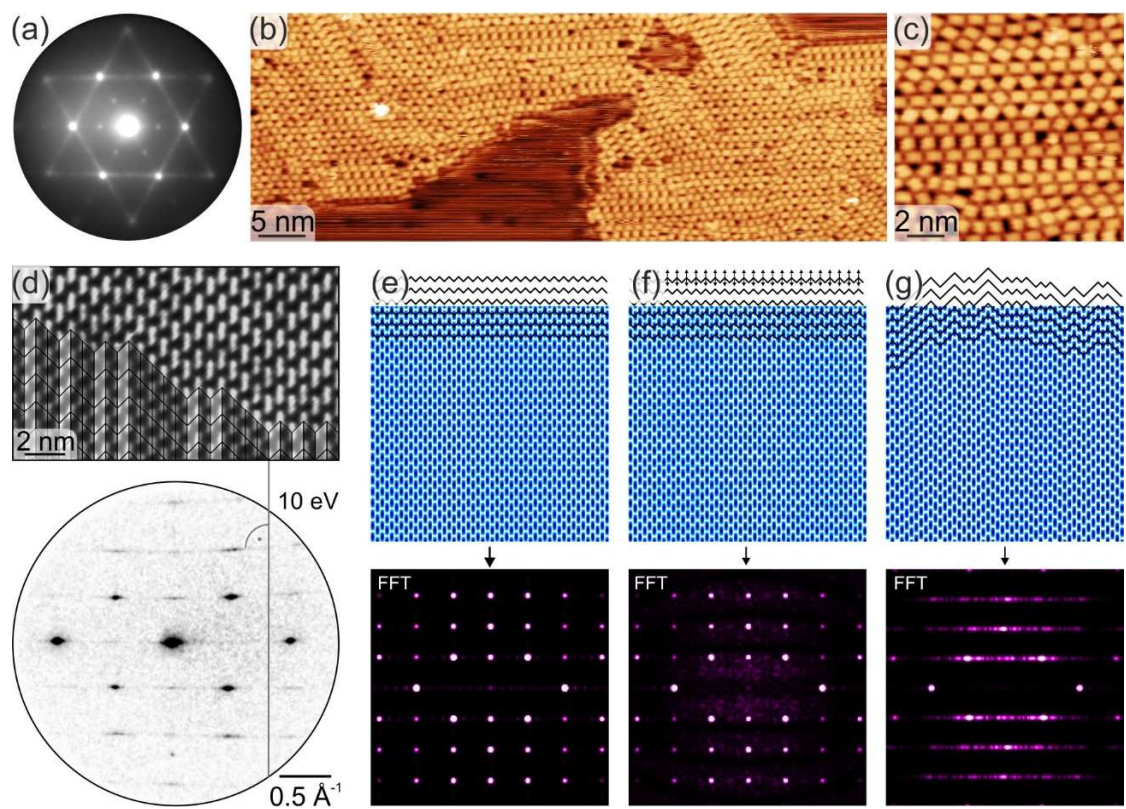

**Figure 3.6.** Line features in diffraction patterns. (a) Diffraction pattern and (b) and (c) STM images of Fe-tetracyanobenzene metal-organic framework on the Au(111). (d) STM image of the BDA α-phase layer with highlighted domain walls and the corresponding microdiffraction pattern given below. The lines highlight the fact that the diffraction streaks (in reciprocal space) are always perpendicular to the direction of domain walls (in real space). (e) – (g) Simulation of random displacement of molecule-like objects and its influence on the resulting FFT signal. (e) The ideal zig-zag arrangement of molecule-like objects results in sharp FFT signals at expected places. (f) Random vertical displacement of all individual objects from the ideal position (by ± 0.5 Å) results

in a decrease of spot intensities but does not affect the positions FFT signals. (g) Random distribution of domain walls leads to the formation of clear streaks in the FFT signal. Parts of the figure were adapted from the ref. [51], Copyright 2025, with permission from Elsevier.

**3.2 Understanding the bright-field images.**

Real-space projection of surfaces with nanometer resolution represents the LEEM fundamental imaging mode. The unrestricted real-space image contains contributions from all backscattered electrons, including those inelastically scattered or diffracted to different angles, which suppresses some of the contrast mechanisms and may increase the signal-to-noise ratio and produce blurred images. By placing the contrast aperture into the back focal plane of the objective lens and collecting only elastically scattered electrons from the central (0 0) diffraction spot, one can significantly improve the image quality and contrast, which is referred to as LEEM bright-field imaging. Bright-field imaging enables the visualization of subtle structural details such as surface reconstructions, atomic steps, magnetic domains, and a number of atomic layers in two-dimensional materials. We note that basic image corrections, i.e., dark-field and flat-field corrections, are necessary for correct quantitative analysis based on bright-field images [52]. While for the diffraction, the formulation of diffraction conditions provided the basic understanding, the understanding of bright field images is based on the electron reflectivity as a function of energy for which no simple theory exists.

The bright-field images are illustrated in Figure 3.7a, which shows the Ag(001) substrate partially covered by the BDA molecules forming two distinct phases, indicated as $\alpha$ and $\beta$. At 10 eV, molecular islands of both phases appear as darker areas and can be easily distinguished from the substrate. At this energy, the brightness of both phases is similar, but they still appear differently due to subtle effects influencing the spatial distribution of the electron reflectivity. The primary advantage of bright-field imaging is the possibility of real-time observation of surface processes during time or as a response to external stimuli like temperature, material deposition, gas exposure, and irradiation. The time resolution allows us to study the dynamics of surface processes, such as molecular phase transformations. This is demonstrated in Figure 3.8, which shows thermally induced BDA α to β phase transformation. Adjustment of the incident electron energy can help distinguish both phases during the transformation, as each phase may exhibit different electron reflectivity at various energies, as discussed below. In this case, real-time bright-field imaging allowed precise characterization of overall transformation dynamics.

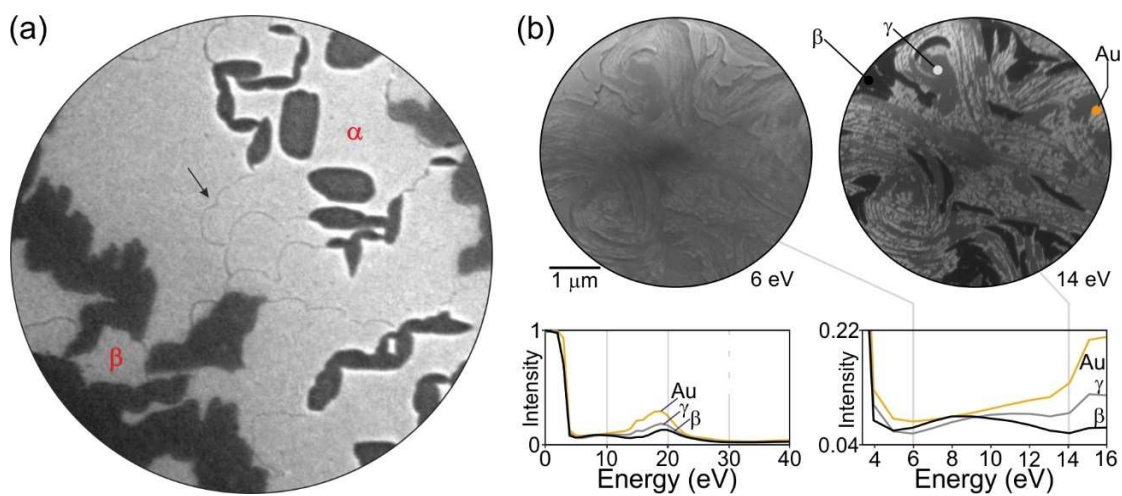

**Figure 3.7.** LEEM bright-field imaging. (a) Bright-field image on Ag(001) surface featuring two distinct BDA phases, α and β. The step edge is marked by an arrow. (b) Bright field images and LEEM-I(V) curves measured for two molecular phases, the β and γ, comprising two distinct products of the intramolecular ring-closure reaction of 1,3,5-tris(7-methyl-α-carbolin-6-yl)benzene on the Au(111) surface. Whereas the Au substrate is always bright, there is a contrast

inversion between β and γ phases at ~9 eV. Parts of the figure were adapted with permission from ref. [44]. Copyright 2025 The Authors, published by Wiley-VCH GmbH under CC BY-NC 4.0 license.

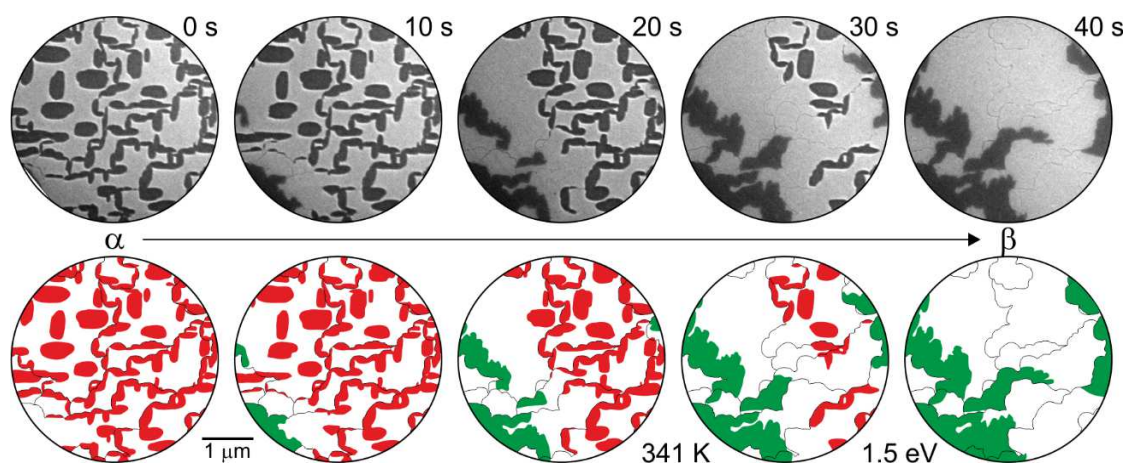

**Figure 3.8.** Real-time bright-field observation of thermally induced BDA α to β phase transformation on the Ag(001). The molecular islands appear as dark areas. The bottom row visualizes the change of spatial arrangement of both phases (α: red, β: green) and the position of step edges (black lines). See video SV1.

### 3.2.1 Contrast mechanisms

Low-energy electrons reflected/scattered from the sample provide several contrast mechanisms in the bright field images [4–7]. Changes in electron reflectivity that occur due to the change in amplitude or phase of reflected electron wave provide contrast between places with different material, structure, or morphology.

*Topographic Contrast*. There is always the topographic contrast caused by surface roughness, inclination of surface elements with respect to the flat surface, or three-dimensional features, which distort the field in front of the object [5,7].

*Mirror Electron Microscopy (MEM)*. For low primary electron energies, the entire electron beam is reflected without reaching the sample surface. In this mode, the low-energy electrons are sensitive to the spatial (temporal) variations in the electric field in the surface plane. As the electrons do not reach the sample, the damage from the electron beam is minimal [6].

*Reflectivity/Amplitude Contrast*. The spatial variation of the amplitude of the electron wave reflected from the surface is the most frequent contrast mechanism in LEEM [4]. The contrast arises from different reflection coefficients of different materials or structures. For example, if two structures have distinct intensities of the diffracted beam, the intensity of the specular (0 0) beam is affected, resulting in the contrast between those two structures [4]. The reflectance also increases at primary electron energies matching the energy gaps of the crystal [5]. Under the single scattering approximation, the reflected intensity would be directly related to the band structure of the surface: the reflectivity shows pronounced maxima at energy regions of band gaps or with low density of states [5,6]. This should be taken as the first approximation, as multiple scattering cannot be neglected at low electron energies.

*Phase Contrast*. Modifications of the phase of the electron wave reflected from the sample give rise to interference effects. For thin film, interference can occur between the electron waves that are reflected from the film surface and the film/substrate interface. The path length difference produces a phase shift between the waves, and consequently, the reflected intensity is modulated as a function of the incident electron energy and the film thickness [4,6]. At step edges, the phase shift between waves reflected on the top and bottom terraces gives rise to an oscillatory interference pattern of the reflected intensity around the step location; hence, in bright field images, the atomic steps are visible at particular electron energies [4,6].

With respect to molecular films on solid surfaces, all the above-mentioned mechanisms act together and typically are not distinguished. The steps are clearly visible on the bare surface, as shown in Figure 3.7a, where they appear as dark lines. By changing the primary electron energy, single steps can be distinguished from step bunches. The LEEM capability to visualize steps presents an important asset with respect to the growth of molecular layers as the step edges may act as diffusion barriers for molecules, can stop the growth of molecular domains, and induce particular orientations of molecular domains.

The molecular layers usually appear darker on the bright substrate. As distinct molecular structures have different reflectivity, they can be distinguished in bright-field images at specific energies. For example, in Figure 3.7b, the reflectivity of particular molecular structures changes with primary electron energy. From measured LEEM-I(V) curves detailed below, we find energies at which the contrast between the phases and substrate is highest (in this case, 14–18 eV); for energies below 9 eV, the contrast reverses, but it is much smaller than in the optimal case.

At higher energies, the contrast between the molecular layers and the substrate becomes lower. Moreover, these "high" energy electron beams strongly affect the molecular layer, often called beam damage. This is visible as darkening of the irradiated areas in the bright field or blurring and weakening diffraction spots with time. The lowest possible primary electron energy should be used to suppress beam damage, and from time to time, the observed area should be changed to ensure that the observed changes in the molecular layer also occur outside the electron beam.

In the bright field, we can also probe the diluted molecular phase, often referred to as 2D molecular gas. The presence of molecular gas induces the darkening of apparently empty surfaces with increasing molecule density.

**3.2.2 LEEM-I(V)**

In the LEEM bright-field imaging, the sample is irradiated by a parallel beam of electrons, and only the (0 0) diffracted beam is selected by the contrast aperture to form the image of the sample surface. The variation of (0 0) spot intensity with primary electron energy, i.e., the I(V) curve, can be recorded for each pixel in the image. This is often referred to as LEEM-I(V) [8]. The LEEM-I(V) contains rich information on the structure of the topmost few atomic layers of the studied material, as discussed with respect to LEED I(V) above. In the case of LEED I(V), many beams are measured in a wide energy range and used to get the sample structure via comparison with the modeled curves. While this is, in principle, possible also for LEEM-I(V) [16], this is not a common practice; instead, LEEM-I(V) is used in a fingerprinting fashion, which allows to distinguish one surface structure from another with high spatial resolution down to level of one pixel [8]. The advanced techniques cover local potentiometry, work function measurements, and layer counting of 2D materials [8].

We will demonstrate fingerprinting on graphite intercalation by organic molecules, where the changes take place in the subsurface region and, therefore, are not directly accessible to local probe measurement. Figure 3.9 shows the bright-field images and associated LEEM-I(V) curves measured for reversible intercalation of graphite by heptanoic acid [53]. After exposure of freshly exfoliated graphite (Figure 3.9a) to heptanoic acid, the bright-field image (Figure 3.9b) clearly changes, but it is restored after vacuum annealing (Figure 3.9c). The measured LEEM-I(V) curves in the energy range of 0–40 eV show clear differences between freshly cleaved and intercalated graphite, especially in the 20–30 eV energy range. The LEEM-I(V) curve obtained after high-temperature annealing again closely resembles that of freshly cleaved graphite except for absolute intensity, indicating the removal of intercalated solvent molecules during this treatment. The second look at heptanoic acid intercalated graphite reveals that the bright-field image shows two distinct levels of brightness (Figure 3.9e), whereas the associated LEEM-I(V) curves (Figure 3.9f) show only a change of intensity between these regions pointing toward minimal structural differences between these regions.

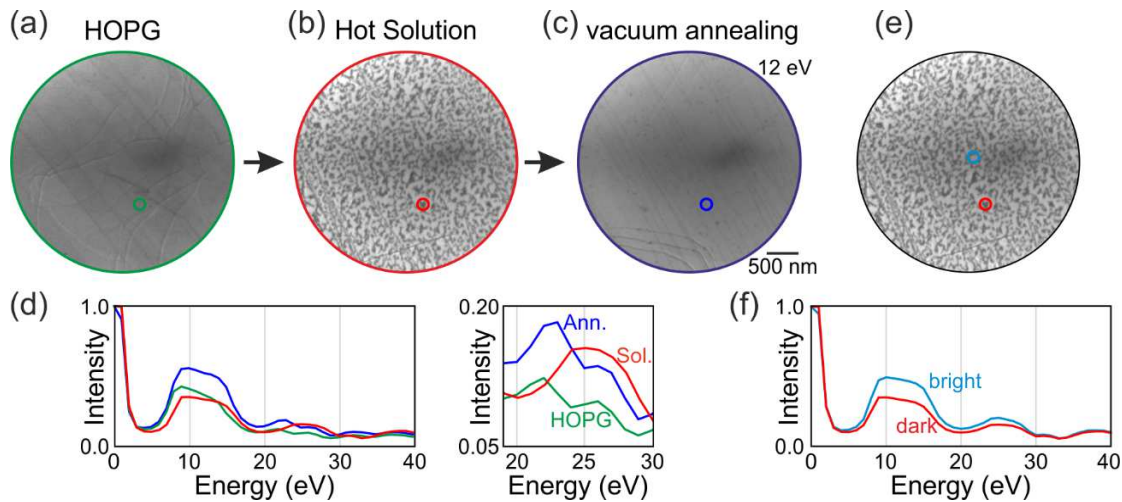

**Figure 3.9.** LEEM-I(V) analysis of intercalation of graphite by heptanoic acid. LEEM bright-field images acquired at 12 eV on (a) freshly cleaved graphite, (b) after immersion in 60 °C hot pure heptanoic acid for 3 days, and (c) after additional ex-situ vacuum annealing at 900 °C for 1 h. (d) LEEM-I(V) curves extracted from the areas marked by the circles in the BF images (colors also match the borders of the corresponding LEEM images); the right-hand side shows the magnification of 20–30 eV region. Both BF images and I(V) curves consistently indicate changes after solvent exposure that are absent in freshly cleaved graphite and removed by vacuum annealing. (e) and (f) For heptanoic acid intercalated graphite, two distinct levels of brightness are observed in bright-field (e), but the associated LEEM-I(V) curve (f) shows only a change of intensity between these regions. Figure adapted with permission from ref. [53]. Copyright 2024 The Authors, published by Wiley-VCH GmbH under CC BY 4.0 license.

**3.3 Dark-field imaging.**

If we select other than the (0 0) diffraction spot by the bright field aperture, we image only electrons contributing to the particular diffraction spot. This mode is, therefore, far more selective than bright field imaging. Only places with a given structure and orientation with respect to the substrate are imaged with high brightness, as demonstrated in Figure 3.10b. Selecting diffraction spots associated with distinct phases and domains (Figure 3.10c) and measuring dark-field images for all of them enables us to make a composite image showing the spatial distribution of individual phases or domains given in Figure 3.10d. This imaging mode is, therefore, a surface science equivalent to phase analysis known from material sciences or electron-backscattered diffraction (EBSD) analysis from classical scanning electron microscopy.

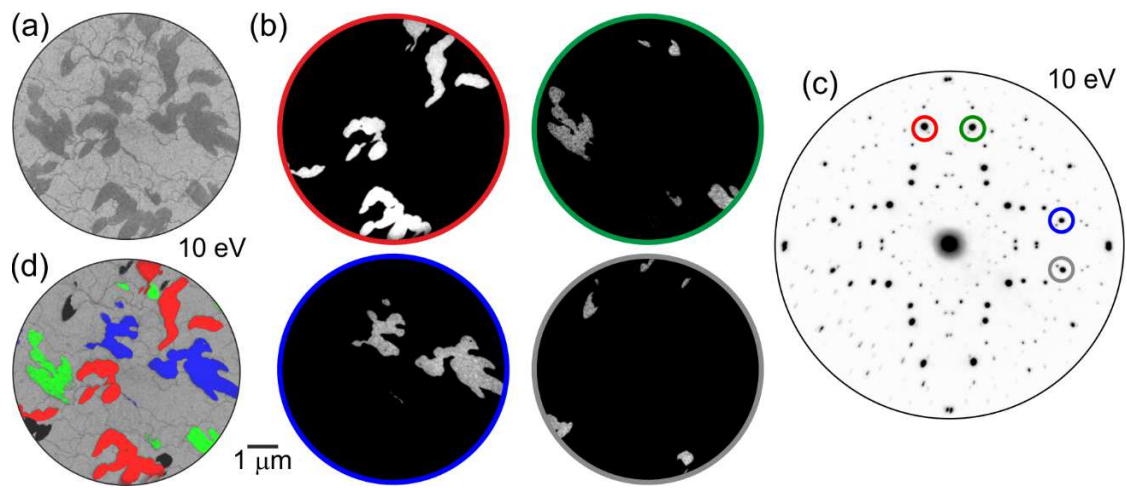

**Figure 3.10.** Dark-field imaging. (a) LEEM bright field image featuring islands of BDA β phase on Ag(001) surface. (b) Four colored dark-field images, each taken with contrast aperture selecting a particular diffraction spot given in (c). Dark-field images show only the islands contributing to the marked diffraction spots, i.e., having specific structure and orientation with respect to the substrate. (d) Color-coded composition of dark field images.

## 3.4 Additional techniques

The LEEM offers several advanced measurement modes, which we will only briefly mention and leave the reader to study the referenced literature. First, the prism also serves as an energy filter, providing the LEEM system the capability to measure angle-resolved energy spectra at down to 250 meV energy resolution and energy-filtered images with a 1.7 eV energy pass band when combined with UV light sources [8,24]. Next, under specific conditions, momentum microscopy measurements are possible in LEEM/PEEM setup, as demonstrated for NTCDA/Cu(001) [54]. Finally, reconstruction of the series of images taken with different values of defocus may be used to improve the resolution, extract the amplitude and phase information for the scattered wave, and remove particular image artifacts [55].

## 4. Case studies

In this section, we will demonstrate the capabilities of LEEM on selected systems that have been well-studied by LEEM. We will focus on self-assembled molecular systems formed by organic molecules. We will not include layers of organic semiconductors like pentacene [56,57] that have been covered separately by Mayer ze Heringdorf [14] and recently in a thesis by Tebyani [58]. Similarly, diindenoperylene and a metal-free single molecular magnet, NitPyn, were reviewed separately [15]. There are also interesting PEEM results for discussed molecular systems [59,60], but they also fall beyond the scope of this review. The self-assembled phases investigated by LEEM include the following molecules: aromatic carboxylic acids [32,41,67–72,43,49,61–66], PTCDA [55,73–76] and NTCDA [54,77,78], phthalocyanines [48,55,73,75,79–81], TCNQ [82–85], and helicenes [86].

### 4.1 Carboxylic acids on Ag and Cu surfaces

The aromatic carboxylic acids, primarily biphenyl-4,4'-dicarboxylic acid (BDA, Figure 4.1a), are one of the most studied molecules forming self-assembled phases employing LEEM. Many examples in the preceding sections were demonstrated on the BDA model system on Cu(001), Ag(001), and Ag(111) substrates. BDA features two phenyl rings that impose its flat-laying geometry on metal surfaces and two carboxylic end-groups mediating intermolecular hydrogen bonds and enabling the formation of extended supramolecular assemblies. These groups can be chemically transformed on substrates: contact with the substrate (Cu) [43,87] or sample annealing to elevated temperatures (Ag) [68,70] leads to dehydrogenation (also called deprotonation, see Figure 4.1b and c), i.e., dissociation of hydrogen from one or both carboxylic groups of BDA.

The BDA deprotonation is significantly influenced by substrate material and surface plane orientation. On Cu(100), BDA forms a self-assembled structure consisting of fully deprotonated molecules already at room temperature [43]. The situation is slightly different on Cu(111): at room temperature, a mixture of partially and fully deprotonated phases appears on the surface [87]. This indicates a lower reactivity of the (111) surface compared to the (100). However, in general, Cu has considerably higher reactivity for deprotonation of BDA than Ag and Au. BDA on Ag(100) deprotonates very slowly at room temperature, and annealing at higher temperatures is necessary to reach a noticeable degree of deprotonation in a reasonable time [70]. In this respect, the Ag(111) surface is less reactive than Ag(100) [70]. No deprotonation of BDA on Au substrates has been observed in our group, which contradicts the LEEM work of the report of Schwarz et al. [66] but is consistent with other experimental reports [88]. Hence, the substrate material has a decisive role in the deprotonation of deposited carboxylic acids, and surface plane orientation also affects the reactivity and structure of molecular phases.

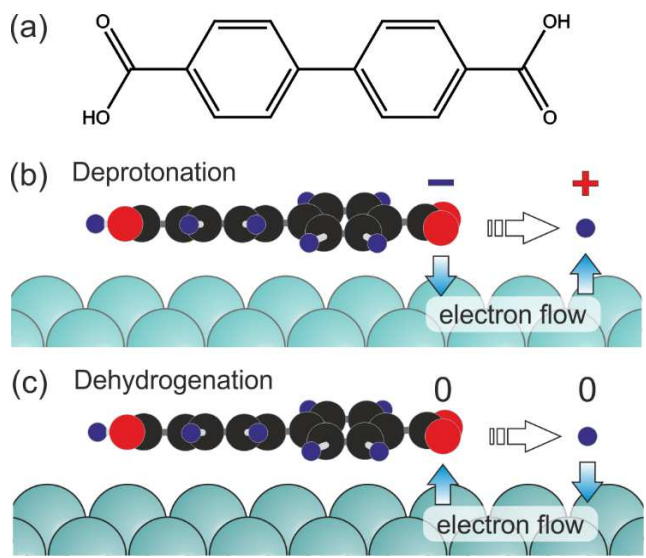

**Figure 4.1.** BDA and on-surface deprotonation or dehydrogenation. (a) Chemical structure of biphenyl-4,4'-dicarboxylic acid. (b) In deprotonation, the removal of a positive proton leaves

negative oxygen atoms. Subsequent charge transfer makes both proton (hydrogen) and oxygen atoms partially charged. (c) Removal of hydrogen leaves neutral oxygen atoms. Subsequent charge transfer makes both hydrogen and oxygen atoms partially charged, i.e., the result is the same as in (b).

**4.1.1 Thermal deprotonation**

The understanding of the deprotonation process and discovering of all the phases would be hardly possible without real time observation of the thermal induced process in LEEM. In following, we will discuss the thermal deprotonation process before explaining particular phases of interest or their transformation. Via the thermal deprotonation process (Figure 4.2a), we obtain a series of molecular phases that comprise a mixture of pristine, semi-deprotonated (one of the carboxyl groups is deprotonated), and fully deprotonated BDA molecules with a given ratio. On Ag(001), the gradual deprotonation of a submonolayer coverage of BDA leads to the formation of multiple molecular phases that we refer to as α, β, $\gamma_3$, $\gamma_2$, and δ presented in STM images given in Figure 4.2b and diffraction patterns in Figure 4.2c [32,41,68]. In the full layer, the ὰ phase is formed instead of the β phase due to the spatial constriction [68]. $\gamma_3$, $\gamma_2$ are intermediate phases formed from the β phase; $\gamma_3$ has the same chemical composition but a different structure, whereas $\gamma_2$ differs both in structure (similar to the $\gamma_3$) and degree of deprotonation. There are two distinct fully deprotonated phases, δ and ε. The δ phase is obtained by annealing above 400 K, whereas the ε phase is obtained only non-thermally, e.g., by e-beam irradiation [69].

The BDA phases named above differ in the degree of deprotonation of carboxyl groups, i.e., α comprises only the pristine BDA with fully protonated groups, β, ὰ, $\gamma_3$ phases comprise 50% of deprotonated groups, $\gamma_2$ phase 66 %, and δ and ε phases 100 % [32,41,68,69]. The deprotonated carboxyl oxygens form chemical bonds with the substrate; the released hydrogen associatively desorbs from the surface, which is driven by a considerable decrease in the free energy of the system [89]. Higher temperatures are necessary to obtain a higher degree of deprotonation. For example, the β phase displays a fixed degree of deprotonation in a broad temperature range, which points to hindered deprotonation within this phase [70]. Stabilizing the degree of deprotonation within condensed phases renders the BDA phases metastable and significantly influences the transformation kinetics. Each of the above-mentioned phases shows interesting features. The α phase represents an ordered phase that is non-periodic. The β phase is a non-glassy disordered self-assembled phase: it is long-range ordered but disordered on a molecular level with a random distribution of pristine, semi-, and fully deprotonated BDA molecules [32]. Finally, γ and δ phases represent *k*-uniform tilings [41].

The degree of deprotonation can be followed by X-ray photoelectron spectroscopy (XPS). Each of the three observed binding motifs has a characteristic spectral fingerprint, given in Figure 4.2d. Complementary hydrogen-bonded carboxyl groups in the α phase manifest themselves by two peaks in the O 1s spectrum; the peaks with an intensity ratio close to 1:1 are associated with hydroxyl (C–O̲H) and carbonyl (C=O̲) oxygens of the carboxyl group as marked in Figure 4.2d [32]. The ἀ and β phases feature the carboxyl-carboxylate binding motif (note that carboxylate is a deprotonated carboxyl group, bottom scheme in Figure 4.2d). In the O 1s spectrum, three peaks with 1:1:2 intensity ratio are associated with hydroxyl and carbonyl oxygen peaks of carboxyl and with two oxygen atoms of carboxylate [32,40]. Finally, the fully deprotonated BDA in the δ phase features a carboxylate-phenyl motif that shows a single peak for both carboxylate oxygen atoms [41,68].

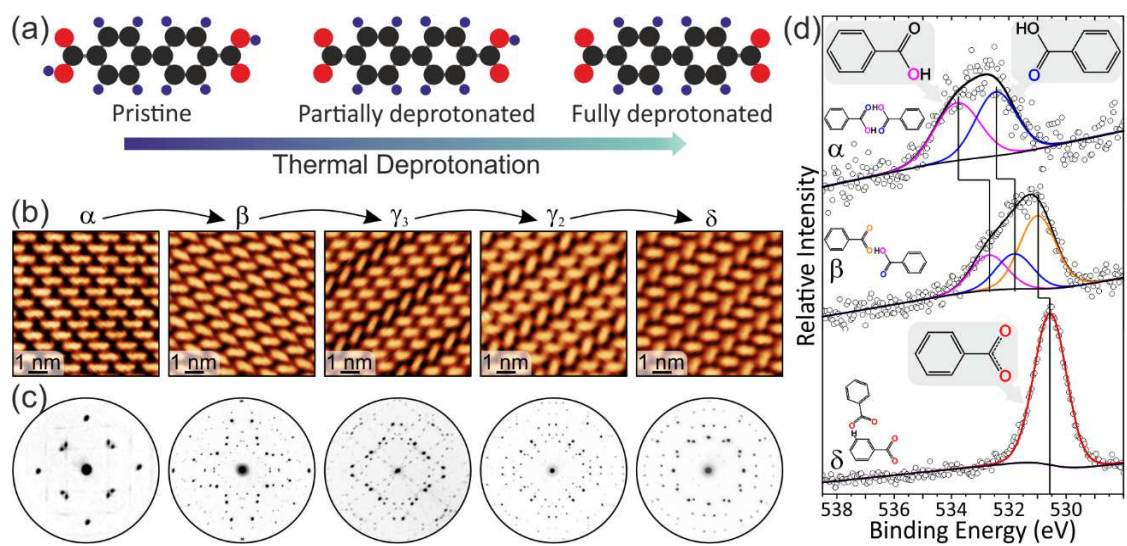

**Figure 4.2.** BDA molecular phases on Ag(001). (a) Schematics of sequential thermal deprotonation of BDA. (b) STM images associated with BDA phases denoted at the top and (c) the associated diffraction patterns. (d) XPS spectra associated with pristine ($\alpha$), partially- ($\beta$), and fully deprotonated ($\delta$) BDA forming particular molecular phases. The oxygen atoms associated with peak components are highlighted by the same color in the gray underlaid schemes. The binding motif for each phase is given on the left side. Parts of the figure were adapted with permission from the ref. [32], Copyright 2020 American Chemical Society; and ref. [40], Copyright 2021, with permission from Elsevier.

## 4.1.2 Phase transformations in real time and space.

BDA forms several molecular phases that appear with its increasing deprotonation. We can follow the transformation of one phase into the subsequent one in real space and real time by employing LEEM. We use the term "transformation" to highlight that the presented phase transitions are irreversible. The irreversibility is a direct consequence of associative hydrogen desorption from the surface under UHV conditions [89]. The reverse process, i.e. dissociative hydrogen adsorption, occurs very rarely due to low hydrogen partial pressure in the surrounding environment. Thermodynamically, this is expressed by a large free energy decrease of 0.5 eV per H atom associated with the entropy of the desorption process [89].

The $\alpha \rightarrow \beta$ transformation is demonstrated in video SV1 (for description, see Figure 4.3a). Nuclei of the $\beta$ phase islands form simultaneously at different locations within the LEEM view field. Once the first nuclei of the $\beta$ phase islands appear, the $\alpha \rightarrow \beta$ transformation is relatively quick. The $\beta$ phase islands grow at the expense of the $\alpha$ phase domains in a manner that we refer to as remote dissolution [32]. Here, an expanding capture zone exists around each $\beta$ phase island, and when this zone reaches the surrounding $\alpha$ phase islands, these quickly dissolve. We have explained this behavior using a general growth-conversion-growth model validated by kinetic Monte Carlo simulations. The transformation follows the La Meer burst nucleation mechanism [90], which we extended to the concept of "burst transformation" [70]. Burst nucleation appears as a consequence of a large critical nucleus size due to relatively weak intermolecular interactions. The molecular phases are in a dynamic equilibrium with the corresponding 2D gas of BDA molecules diffusing over the surface. The 2D gas plays an indispensable role in phase transformations: it mediates mass transport between the individual islands and hosts the deprotonation reaction [32]. While deprotonation also occurs within the condensed phase, the intermolecular bonds stabilize the initial state, which increases the activation energy for deprotonation. Therefore, the deprotonation is much faster in 2D gas than in the condensed phase.

The general picture of burst transformation is as follows. In the beginning, there are islands of the initial phase in equilibrium with 2D molecular gas. With the steady temperature increase, more molecules are released to the 2D gas, where they can deprotonate. This gradually builds up the 2D molecular gas, and continuing deprotonation makes its composition different from the parent phase. The equilibrium condition requires more molecules to be detached and eventually deprotonated. At some point, sufficient supersaturation allows the formation of the new phase – its formation follows the La Meer burst nucleation mechanism ([Figure 4.3c](#)). The presence of new islands rapidly decreases supersaturation: the molecules from the 2D gas are attached to them. This has two effects: (i) without significant supersaturation, no additional nucleation occurs, and (ii) many molecules are released from the preceding phase, rapidly deprotonate, and are incorporated into new phase islands. This process, i.e., remote dissolution, continues until the islands of the initial phase are dissolved [70].

We have observed different kinds of phase transformations between BDA phases on Ag(001). The $\alpha \rightarrow \beta$ transformation can be considered a first-order transition [32]. The structure of the growing phase ($\beta$) is incompatible with that of the previous phase ($\alpha$). Therefore, the islands of the $\beta$ phase nucleate outside the $\alpha$ islands. The situation is different for transformations between the $\gamma$ phases. The structure of the $\gamma$ phases shares the structure of one side of their unit cells. Hence, a new phase can grow on the periphery of the previous one. The transformation between the $\gamma$ phases then proceeds locally within the voids propagating through the molecular islands or at the island periphery; however, the overall shape and position of the islands are maintained [41], see video

SV2, and description in Figure 4.3b. In contrast, the α → ά transformation is gradual and proceeds without transport of molecules [68]; hence, can be termed a second-order irreversible transition.

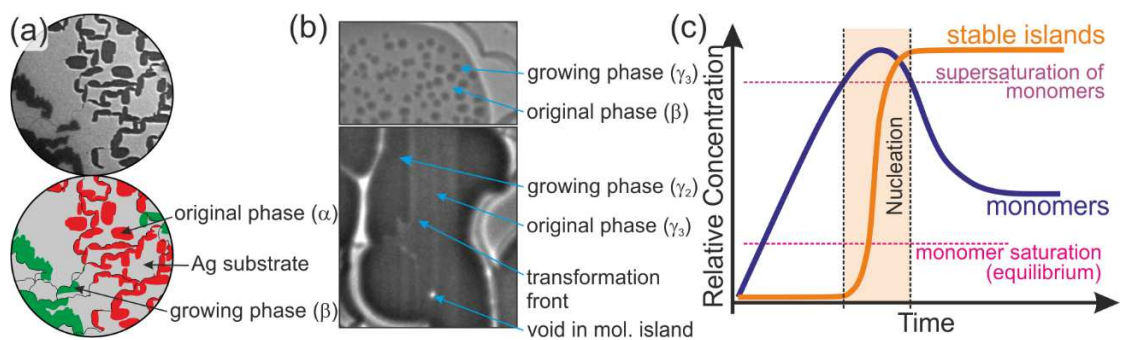

**Figure 4.3.** Phase transformations on Ag(001). (a) Graphical description of Video SV1 showing the remote dissolution of the α phase and its transformation into the β phase. (b) Graphical description of Video SV2 showing first the β → $γ_3$ transformation followed by $γ_3$ → $γ_2$ transformation. The latter proceeds internally, i.e., within existing molecular islands. (c) Burst nucleation process. A significant supersaturation has to be reached before the nucleation starts. Once reached, the nucleation events appear quasi-simultaneously. When the stable islands are present, the precursor molecules attach to them, decreasing the supersaturation below the point where nucleation can occur. Consequently, only the existing islands grow, and no further nucleation is possible. Parts of the figure were adapted with permission from reference [41], Copyright 2019 The Authors, published by Springer Nature under CC BY license; and from reference [90], Copyright 2015 Royal Society of Chemistry.

### 4.1.3 The complex structure and diffraction pattern of the α phase

Atomic and molecular phases usually show long-range order unless they are amorphous. BDA molecular phases do not differ from this concept except for the α phase. The α phase is aperiodic in one direction, features a high concentration of twin domain boundaries as an inherent feature, and the associated diffraction pattern is challenging to understand.

At sub-monolayer coverage, the α phase forms needle-like molecular islands oriented along the primary substrate crystallographic directions (Figure 4.4a). The needle-like morphology reflects the internal structure of the islands. The BDA α phase structure consists of extended molecular chains oriented along the long axis of the islands (Figure 4.4b) [32,68]. Within the chains, the protonated carboxyl groups of neighboring BDA form complementary hydrogen bonds. The BDA chains are connected by comparatively weaker hydrogen bonding; cf. 0.9 eV per molecule within the chain with 0.3 eV in between the chains [51]. Within the chains, BDA molecules are not periodically spaced, but they are shifted from "optimal" positions by ± 0.4 Å to balance molecule-molecule and molecule-substrate interactions [40], as discussed for a relevant system comprising non-deprotonated terephthalic acid on a Cu substrate [91]. The key structural motif is the pair of BDA molecules in the neighboring chains, i.e., BDA dimers marked in Figures 4.4b and f; the dimeric motif spatially extends over many molecules [68]. The mutual position of BDA molecules within the dimer comes with two equivalent possibilities; the switching between these orientations appears as a twin domain boundary, marked in Figure 3.6d.

The μ-diffraction and diffraction patterns shown in Figure 4.4c–e comprise several features discussed already in Section 3.1.4, i.e., the non-periodicity of the α-phase along the molecular chains resulting only in the weakening of diffraction spots and the diffraction streaks appearing due to presence of twin domain boundaries. We see that the presence of both domains in Figure 4.4d makes the pattern more complex but, on the other hand, easier for the quantitative measurements by the local congruence approach. The reciprocal and real space unit cells are marked by blue rectangles. From the symmetric position of the (1 1) spot with respect to (0 0) and (2 0) spots, we infer that the basic structure of the α-phase consists of the alternating molecular chains as depicted in Figure 4.4f with the associated wallpaper group of $p2mg$. In the diffraction pattern (Figure 4.4d and e), only the (1 1) spot is visible; the (1 0) and (0 1) spots are missing. In the horizontal direction, we identify a glide symmetry axis (highlighted in Figure 4.4g), and, consequently, the odd, i.e., (1 0), (3 0),…, spots should have zero intensity at all energies. However, the absence of (0 1) spot has a different reason: at lower energies, the spot is clearly visible, but it is weak and smeared, as demonstrated in Figure 4.4c. The weakening and smearing of the spot are due to the presence of the twin domain boundaries, as evident from the FFT simulation given in Figure 3.6g.

Upon deprotonation at the full monolayer coverage, the ά phase is formed. It is structurally close to the α phase, but the deprotonated BDA molecules display stronger binding to the substrate via carboxylate oxygen atoms [68]. The stronger BDA-substrate binding results in better-defined intrachain periodicity, which is expressed as much brighter dot features around the main diffraction spots [68]. On the other Ag facet, Ag(111), the α phase can incorporate up to 30 % of deprotonated carboxylate groups, which makes the phase structurally better defined, as documented by the associated sharp diffraction pattern [70].

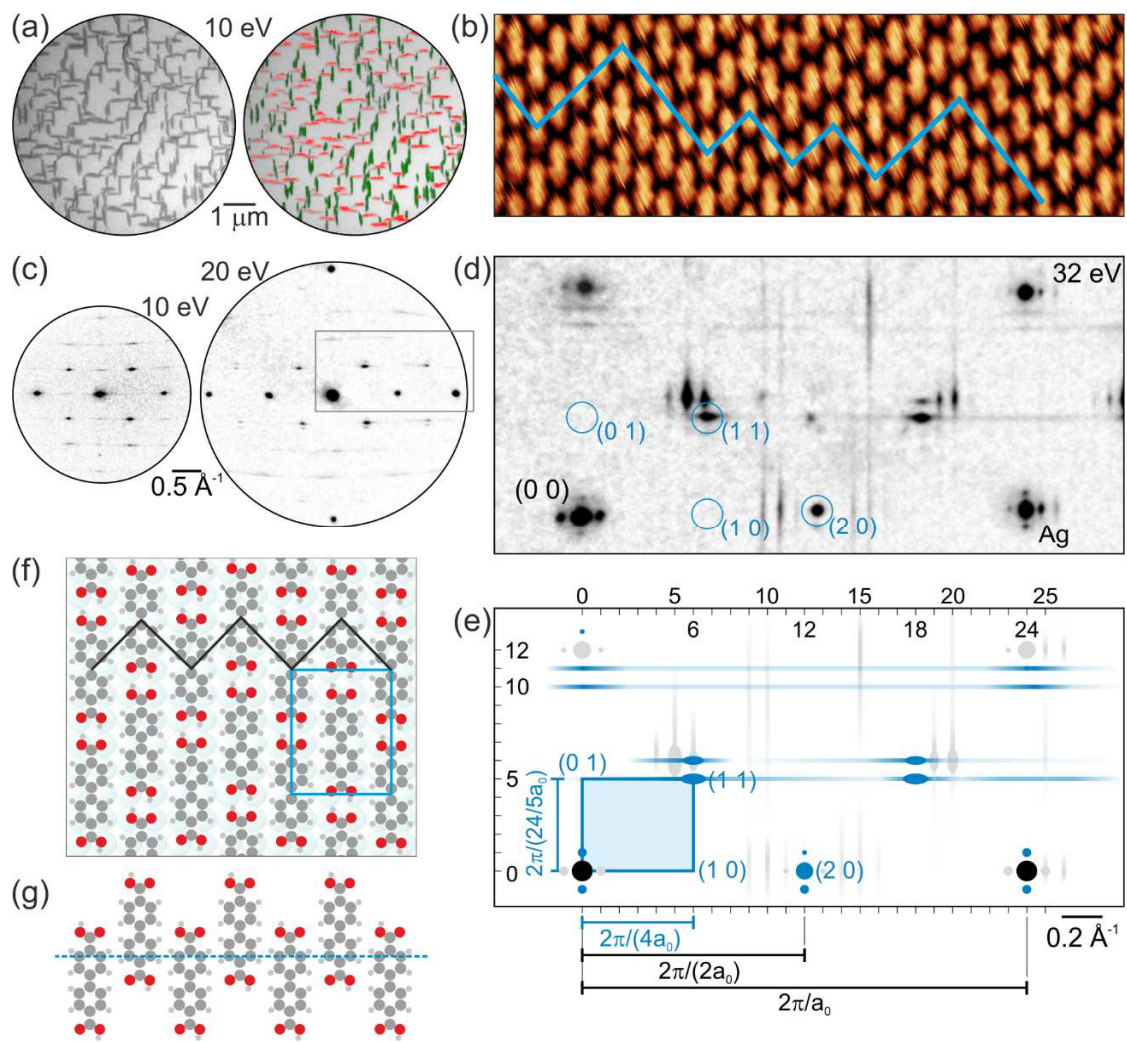

**Figure 4.4.** The BDA α phase on Ag(001). (a) Bright-field image and composition of dark-field images of submonolayer coverage of the α phase. The BDA islands appear as dark (red/green, respectively) protrusions on the brighter substrate. Due to the *p4mm* symmetry of the Ag(001), there are two symmetry-equivalent domains mutually rotated by 90°. (b) STM of the α phase. The molecular chains are oriented vertically. The cyan line connects the centers of BDA molecules, highlighting the BDA dimer motif that spans many neighboring molecules. (c) μ-diffraction patterns measured at 10 and 20 eV energy. (d) Cutout from the full diffraction pattern measured at 32 eV at tilted geometry; its position is marked in the 20-eV pattern in (c). The central spot is marked (0 0), and the molecular layer spots are indexed accordingly. The circles mark missing spots. (e) Modelled diffraction pattern. The substrate spots are black, and the features associated with two distinct α phase domains are blue and gray, respectively. (f) Model of the idealized α phase: C is gray, O red, and H small gray. The blue rectangle marks the unit cell; the black line connects the centers of BDA molecules that form the dimers. (g) The glide symmetry axis associated with the α phase. Parts of the figure were adapted from ref. [40], Copyright 2021, with permission from Elsevier and ref. [51], Copyright 2025, with permission from Elsevier.

**4.1.4 BDA on Cu**

The BDA on Cu was already introduced to give an example of glide symmetry in Section 3.1.4 and Figure 3.4. BDA on Cu(001) is already fully deprotonated at room temperature; hence, only one molecular phase is present for sub-monolayer coverage. Following the time evolution of contrast in the bright field provided a deep knowledge about the nucleation and growth of molecular phases. On Cu(001), the BDA domains nucleate randomly, mostly on the terraces, indicating that the typical heteronucleation on step edges does not play a dominant role here [61]. The follow-up study revealed that steps are permeable to individual molecules but completely impermeable to condensed BDA domains [64]. This was later explained with the help of STM by passivation of the step edges by BDA molecules that prevent nucleation there [43]. The BDA phase at Cu(001) is strained at room temperature. This is expressed in the delayed burst nucleation of the islands that grow only to a certain maximum size, followed by a second round of nucleation events [61]. The growth experiment at higher temperatures shows only a single nucleation phase, as the stress is largely reduced due to the distinct thermal expansion of the layer and substrate [61]. Figure 4.5a shows that at 332 K, the nucleation occurs at coverages of 0.030 ML; after nucleation, the concentration of BDA in molecular gas decreases to an equilibrium concentration of 0.024 ML [62] following the curve given in Figure 4.3c. Bright-field imaging enables the observation of the formation, decay, and size fluctuation of subcritical islands, which allows us to extract the local chemical potential and quantitative thermodynamic description [62]. With increasing temperature, the concentration of BDA in 2D molecular gas in equilibrium with the condensed phase islands (Figure 4.5b) increases from 0.05, 0.10 ML, to 0.3 ML (1ML is fully covered surface) at temperatures 340, 370, and 420 K, respectively [63]; the ratio on BDA being in the molecular gas and in condensed islands increases, which was used to determine the free energy per molecule in the condensed phase. During the deposition at elevated temperatures, a compressed BDA phase nucleates in the center of large (uncompressed) domains, but after turning the deposition off, it decays to the original BDA phase [65].

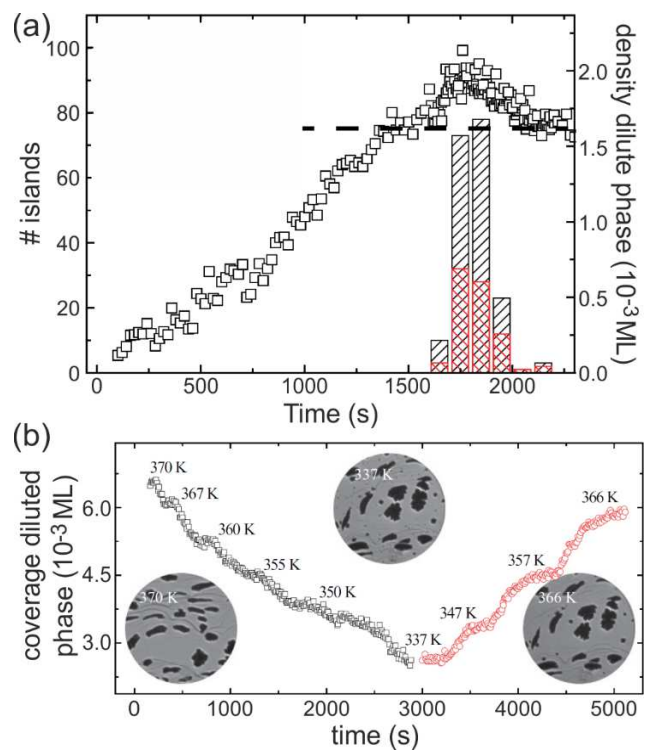

**Figure 4.5.** The BDA on Cu(001). (a) Number of nucleating and decaying islands in 12.5 μm² area (bars, left axis) and density in the diluted phase (i.e., 2D molecular gas). Here, the full coverage by molecular phase is 0.0625 ML, as it refers to the concentration of substrate atoms. The red part of the bars represents the number of decayed islands. The deposition starts at 0 s; the dashed horizontal line marks the equilibrium BDA density. (b) Experiment to determine the equilibrium BDA density in the diluted phase from the brightness of the area surrounding the islands. The BDA islands were grown at 370 K, and the sample was progressively cooled and subsequently heated again. Figure (a) reprinted with permission from reference [62], copyright 2012 by the American Physical Society. Figure (b) reprinted with permission from reference [63], copyright 2012 by the American Physical Society.

## 4.2 PTCDA and NTCDA

Perylene derivative 3,4,9,10-perylene-tetracarboxylic-dianhydride (PTCDA) [55,73–76] and its naphthalene counterpart 1,4,5,8-naphtalene-tetracarboxylic-dianhydride (NTCDA) [54,77,78] are amongst the most intensively studied large π-conjugated molecules on surfaces by LEEM.

Growth of PTCDA on Ag(111) follows the mechanism described above for BDA: after a delayed onset, nucleation occurs at defects, i.e., steps, step bunches, or point defects; once finished, no new nucleation is observed [76]. The initial nuclei are a few micrometers apart, pointing to facile diffusion of PTCDA molecules on the Ag(111) surface. Step bunches were found to stop the growth of islands completely; on the contrary, monoatomic steps just hinder the growth of islands, but they can be overgrown by an expanding island [76]. The PTCDA on Ag(111) is known for hosting many substrate-symmetry-equivalent domains (rotational and mirror) as well as translational domains. All these domains are equally probable, and small and medium terraces usually host only one domain, whereas different rotational/mirror domains may coexist only on significantly large terraces. The orientation of domains is not random; in most cases, there is a strong correlation between step direction and domain orientation [76]. In the second layer, the onset of delayed nucleation is 2−3 longer than for the first layer, and the distances between the nucleation centers are significantly larger as well; however, this is not translated to the domain sizes, which are restricted by domain boundaries of the first layer [76]. Similar growth behavior is observed with NTCDA, which has the same symmetry and functional groups but a smaller aromatic core (naphthalene instead of perylene). The most striking difference compared to PTCDA is the vastly increased onset of nucleation of the first layer islands. The necessary supersaturation of molecular gas (visible through the darkening of the substrate) increases with temperature, reaching about 50 % for 365 K [77].

On Au(111), PTCDA experiences much weaker bonding compared to the Ag substrate; however, the growth is very similar for both surfaces. The most significant difference is the increased diffusion barrier on Au(111) compared with Ag(111), which results in higher island densities at the same temperature [74]. On the contrary, NTCDA/Cu(001) grows in dendrite-like, fractal structures [78]. The dendritic structures arise from a preferential growth in one direction, in which a long-range ordered network of thin molecular chains spans over the entire surface already at small coverages; only later, the voids in the network structure are incrementally filled [78].

## 4.3 Bicomponent molecular systems.

### 4.3.1 Mixed molecular phases

LEEM is an ideal tool to probe the properties of complex systems, e.g., multicomponent molecular phases, where the complex interplay between intermolecular and molecule–substrate interactions define the structure of resulting phases. In this respect, the BDA phases discussed in Section 4.1 comprise three structurally similar but chemically distinct molecules. In this sense, these can be considered as multicomponent molecular phases.

Systems comprising copper-phthalocyanine (CuPc) and PTCDA molecules are particularly interesting from both fundamental and application points of view [73,75]. CuPc molecules on Ag(111) show a weak intermolecular repulsion and, therefore, cover the entire surface in the form of diluted 2D molecular gas. PTCDA exhibits a permanent electrostatic quadrupole moment and forms compact islands on Ag(111). When mixed together, CuPc and PTCDA form three distinct mixed phases with 1:2, 1:1, and 2:1 ratios of the parent molecules, which were structurally detailed by the combination of SPA-LEED and STM with the help of DFT calculations [73]. The condensed phase always coexists with the 2D gas phase, which acts as a reservoir of CuPc molecules. The density of the molecular gas is the decisive parameter for the phase formation in the molecular layer; increasing the gas phase density leads to the formation of more CuPc-rich crystalline structures [75].

The formation of the mixed phases followed by LEEM shows distinct kinetics depending on the deposition order. Starting with a pre-coverage of CuPc, which is present on the surface as molecular gas, leads to the immediate formation of the thermodynamically most stable structures upon deposition of PTCDA at room temperature [73]. The change of deposition sequence, i.e., starting with PTCDA islands on the surface and subsequent deposition of CuPc at room temperature, results in the co-existence of PTCDA islands and CuPc molecular gas; the formation of a mixed structure can be achieved either by post-annealing the sample at ~575 K or by CuPc deposition at an elevated temperature of 375 K [73].

Considering the other systems comprising phthalocyanines studied by LEEM, we have already mentioned the $F_{16}$CuPc–CoPc mixture to discuss the influence of the molecular structure factor on the intensity of diffraction spots in chiral diffraction patterns [48]. Another system studied by LEEM features a mixture of CuPc and $C_{60}$ molecules on Si(001), where the formation of 10 nm-sized domains was observed for a 1:1 mixing ratio [80].

Formation of the heteromolecular mixed phases can be limited to submonolayer coverages of the constituents. BTB molecules introduced in Section 5.1.4 form robust layers on both Ag(111) and Ag(100) surfaces. If these layers cover the entire substrate, they do not mix with organic semiconductors, e.g., pentacene, placed on top. However, if BTB is co-deposited with pentacene, mixed pentacene–BTB phases are formed, indicating their thermodynamic preference [72]. In both cases, the decisive factor that determines stability is the adsorption energy of a molecule per unit area. There are two main contributions that decrease the free energy of the system: molecule-substrate bonding and intermolecular bonding. The computations reveal that the BTB layer has lower free energy per unit area than the mixed phase, mainly due to the strong O-Ag bonding of BTB over the physisorbed pentacene [72]. This is decisive for the stability of the full monolayer of BTB. For submonolayer coverages, there is a free substrate to accommodate all the adsorbed molecules irrespective of their bonding strength to the substrate; hence, the intermolecular bonding favors the mixed pentacene–BTB phases [72].

### 4.3.2 Metal-Organic Frameworks

Two-dimensional metal-organic frameworks comprise organic molecules bound to metal atoms. LEEM can be used to follow their formation and to reveal the complete set of phases that appear on the surface depending on the ratio of metal atoms to organic linkers, deposition/annealing temperature, or time. Thus, it is quite efficient for finding the parameters for their synthesis. So far, only TCNQ-based networks have been studied by LEEM [82–85]. The LEEM was used to follow the TCNQ deposition [83], identify the transient phases [83], and monitor the incorporation of coadsorbed potassium atoms [82] on Ag surfaces. Separate studies identified up to 15 rotational domains of Metal-TCNQ frameworks (Metal: Fe, Ni, Mn) on graphene/Ir(111) and Au(111); among them, only 2–3 present a unique relationship with the substrate [84,85].

### 4.4 Monitoring chemical reactions

LEEM allows us to perform measurements in conditions at which the chemical reaction occurs, i.e., at elevated temperatures or during gas exposure or deposition of molecules. The application of LEEM, either alone or in combination with spectroscopic photoemission mode, to study reactions of small organic molecules on metal surfaces with respect to heterogeneous catalysis was reviewed recently [7,19,20]. Concerning the large organic molecules, there are two principal ways to look at chemical reactions: via the change in contrast in the bright field (fingerprinting) or by the emergence of specific product phases. Below, we will discuss several examples of these approaches.

**4.4.1 Electron-induced reactions: BDA deprotonation.**

Temperature has a prominent position in natural sciences: many reactions and processes are thermally activated following the Arrhenius law. In self-assembly, thermodynamics usually defines the equilibrium structure. However, the kinetics of the involved on-surface processes, i.e., reaction rate, on-surface diffusion, nucleation, and growth, also play a crucial role. The kinetic rates are thermally activated and display a particular hierarchy of activation energies. The correct choice of temperature enables the adjustment of the rate of structural and chemical transformations to reach the desired product [92]. The non-thermal activation methods present an alternative that circumvents the limitation of the given hierarchy of activation energies. Photons and low-energy electrons can excite specific vibrational modes, access many reaction channels, deliver much higher energies, and excite plasmons for plasmon-induced chemical reactions. Electron-beam-induced reactions such as crosslinking/breaking polymer chains in electron beam lithography [93], or precursor decomposition in electron-beam-induced deposition (EBID) [94] are the key technologies for nanofabrication. The commonly used high-energy electron beams deliver too high energy densities, causing extensive damage to adsorbed molecular layers, substantially reducing reaction selectivity; even low-energy (< 20 eV) electrons are still considered rather more damaging than useful, especially for organic adsorbates [95].

Considering the self-assembled molecular layer on surfaces, LEEM usually has a considerable effect on them, especially at energies necessary for measuring diffraction patterns. Returning to our model system, BDA on Ag, the e-beam irradiation of the α phase leads to the formation of new molecular phases given in Figure 4.6a [69]. Depending on the electron energy, the ε phase is either formed directly or via the intermediate β phase. Whereas the formation of the β phase can be induced by both irradiation and thermal annealing above 340 K, the ε phase is only reached by e-beam irradiation. To determine the deprotonation rate, we have employed the α → $\dot{α}$ transformation that can be induced both thermally and by e-beam; importantly, it proceeds without the transport of molecules, so the effects of diffusion and nucleation are eliminated. The intensity of diffraction spots associated with the $\dot{α}$ phase can be, therefore, used to measure the degree of deprotonation and deprotonation rate (Figure 4.6b). Figure 4.6c shows that the growth rate of the $\dot{α}$ phase increases with an e-beam energy above the threshold of 6 eV.

The growth of distinct phases depending on the e-beam energy is a direct consequence of the value of the deprotonation rate of BDA. Depending on the competition of the deprotonation with the nucleation rate, either the intermediate β phase or directly the final ε phase is formed. Regulation of the deprotonation rate thus allows a change of the limiting factor from the rate of supply of the deprotonated molecules to nucleation of the molecular phase, which results in different products. The main interaction mechanisms of low-energy electrons with molecular layers are non-resonant electron impact ionization and electron impact excitation or resonant electron attachment [96]. As we do not see any clear signatures of resonance peaks, we associate the monotonic increase of the reaction rate above 6 eV with the non-resonant electron impact excitation process.

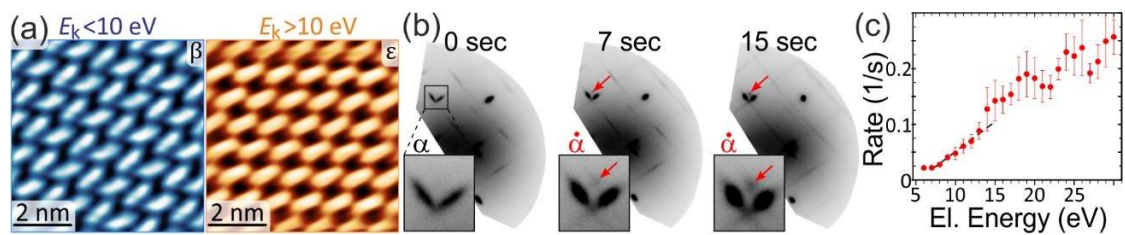

**Figure 4.6.** The electron beam induced deprotonation of the BDA α phase on Ag(001). (a) The electron beam irradiation of BDA/Ag(001) leads to the formation of new molecular phases. Below 10 eV, the β phase grows first, followed by the ε phase. Above 10 eV, the ε phase grows directly. (b) To follow the deprotonation, the internal transformation of the α to ὰ phase was used: the intensity of spots associated with the ὰ phase provides the fraction of deprotonated carboxyl groups. (c) The transformation rate as a function of primary electron energy. Parts of the figure were adapted from the ref. [69], Copyright 2022, with permission from Elsevier.

**4.4.2 Other reactions**

Recently, LEEM was used to follow the intramolecular ring-closure reaction of 1,3,5-tris(7-methyl-α-carbolin-6-yl)benzene on the Au(111), which includes the organization of the precursor to the extended molecular phase, its subsequent dissolution into 2D molecular gas where the reaction took place, and the formation of product phases upon cooling [44]. LEEM provided insights in every step and addressed whether the reaction occurs in the condensed or the 2D gas phase, which presents a critical entry point for theoretical assessment of the reaction mechanism. Moreover, LEEM allowed detailed characterization of both reactants and product self-assembled phases and reliable estimates of product yields.

The second example concerns the crosslinking of self-assembled monolayers of 4'-nitro-1,1'-biphenyl-4-thiol into molecular nanosheets employing the 2.5–100 eV electrons [97]. The cross-linking rate changes by four orders of magnitude in this energy range. At high electron energies (6.5–100 eV), the cross-linking is dominated by the direct electron impact ionization, whereas at low electron energies (2.5–6.5 eV), the dissociative electron attachment prevails.

## 5. Concluding remarks.

LEEM is a UHV surface science technique based on imaging the spatial distribution of a low-energy electron wave reflected from the sample surface. LEEM is a fast technique that provides the first information on the sample in order of minutes after its insertion into the microscope. In dynamic mode, video-rate imaging during sample annealing, irradiation, deposition of new material, or exposure to a specific gas provides information on the kinetics and dynamics of the involved processes.

This review was focused on self-assembled molecular systems on solid surfaces. The case studies highlighted areas where the LEEM is particularly strong or, in several cases, even irreplaceable by other surface analysis methods. LEEM provides *a mesoscale view,* enabling the overview of the representative area of the sample surface and the identification of new and minor molecular phases [82,83]. It enables the analysis of places where the local probe measurements are hardly doable, like step bunches and morphological defects, which STM usually avoids. LEEM *real-time view* enables monitoring the changes during deposition [61,63,77,86], annealing, gas exposure, or other modifications. Thus, it shows delayed nucleation [74,77,79], evolution of molecular phases with deposition [73,81,98], nucleation and decay of sub-critical islands [62], formation of transient phases [65], influence of step edges [43,64] or other morphological defects like wrinkles [67] or distinct orientation of graphene with respect to metal substrate [79], and changes in density of 2D molecular gas [63,77]. During the annealing, it visualizes changes in the structure, e.g., changes in the size of islands [32,63], ordering as a response to the unification of the molecular structure [44], or chemical changes within the molecules [32,40,41,70]. LEEM is ideally suited to assess the effect of electron beams, inducing specific chemical changes and, at the same time, monitoring them [69,97]. Measured *diffraction patterns* help in the determination of unit cells with high precision [32,40,41,70–72], identification of large commensurate unit cells [32], multiple orientations of molecular layers with respect to the substrate [84], chiral structures [44,48], and domain boundaries [49,51]. LEEM *phase imaging* in the dark-field mode shows the spatial distribution of distinct phases on surfaces and enables assessing their mutual correlations [76], e.g., matching with substrate step orientation [76] or constructing phase diagrams for complex, e.g., binary systems [75]. *LEEM-I(V) fingerprinting* enables the identification and distinguishing of molecular phases with high precision or their occurrence in subsurface regions of the substrate [53]. The full potential of LEEM is only reached in combination with other methods of surface analysis, which is documented in the more recent studies in which the LEEM is rarely used alone. LEEM is ideally suited for providing quantitative kinetic and thermodynamic data for molecular systems, enabling us to formulate/refine general theories to predict self-assembly from thermodynamic and kinetic perspectives.

In view of its capabilities, LEEM is surprisingly underrepresented in surface science and related disciplines. One of the reasons can potentially be that it is primarily sold as a standalone instrument with a price tag surpassing the low-temperature STM systems. We believe that LEEM can become a standard tool of surface science if it is attached to the common systems featuring STM and XPS, largely enhancing the rate of measurements by avoiding a series of blind treatments, e.g., annealing/measurement cycles.

AUTHOR INFORMATION

**Notes**

The authors declare no competing financial interests.

ACKNOWLEDGMENT

This research has been supported by GAČR, project No. 25-15999S. We thank Veronika Stará, Anna Kurowská, and Daniela Hrubá for the critical reading of the manuscript.